\documentclass[a4paper,fleqn,usenatbib,useAMS]{mnras}

\usepackage{graphicx}
\usepackage{epsfig}
\usepackage{amsmath}
\usepackage{amstext}
\usepackage{amssymb}
\usepackage{upgreek}
\usepackage{stfloats}
\usepackage{epsf}
\usepackage{newtxtext,newtxmath}
\usepackage{tablefootnote}
\usepackage[T1]{fontenc}
\usepackage{fixltx2e}
\usepackage{ae,aecompl}
\usepackage{afterpage}

\title[SNe 2013fs \& 2013fr]{SN~2013fs and SN~2013fr: Exploring the circumstellar-material diversity in Type II supernovae}

\author[Bullivant et al.]{Christopher~Bullivant$^{1}$\thanks{Email: cwbullivant@email.arizona.edu}, Nathan~Smith$^{1}$, G.~Grant~Williams$^{1,2}$, Jon~C.~Mauerhan$^{3}$, \and
  Jennifer~E.~Andrews$^{1}$, Wen-Fai~Fong$^{1}$, Christopher~Bilinski$^{1}$, Charles~D.~Kilpatrick$^{6}$, \and
  Peter~A.~Milne$^{1}$, Ori~D.~Fox$^{4}$, S.~Bradley~Cenko$^{5}$, Alexei~V.~Filippenko$^{3}$, \and
  WeiKang~Zheng$^{3}$, Patrick~L.~Kelly$^{3}$, Kelsey~I.~Clubb$^{3}$
\\
$^{1}$Steward Observatory, University of Arizona, 933 N. Cherry Ave., Tucson, AZ 85721, USA\\
$^{2}$MMT Observatory, P.O. Box 210065, University of Arizona, Tucson, AZ 85721-0065, USA \\
$^{3}$Department of Astronomy, University of California, Berkeley, CA 94720-3411, USA \\
$^{4}$Space Telescope Science Institute, 3700 San Martin Dr., Baltimore, MD 21218, USA \\
$^{5}$Astrophysics Science Division, NASA Goddard Space Flight Center, Greenbelt, MD 20771, USA \\
$^{6}$Department of Astronomy and Astrophysics, University of California, Santa Cruz, CA 95064, USA}

\date{Accepted XXX. Received YYY; in original form ZZZ}

\pubyear{2017}

\begin{document}
\label{firstpage}
\pagerange{\pageref{firstpage}--\pageref{lastpage}}
\maketitle

%% Abstract
\begin{abstract}
  We present photometry and spectroscopy of SN~2013fs and SN~2013fr in the first $\sim100$ days post-explosion. Both objects showed transient, relatively narrow H$\alpha$
  emission lines characteristic of SNe~IIn, but later resembled normal SNe~II-P or SNe~II-L, indicative of fleeting interaction with circumstellar material (CSM). SN~2013fs was
  discovered within 8~hr of explosion; one of the earliest SNe discovered thus far. Its light curve exhibits a plateau, with spectra revealing strong CSM interaction at
  early times. It is a less luminous version of the transitional SN~IIn PTF11iqb, further demonstrating a continuum of CSM interaction intensity between SNe~II-P
  and SNe~IIn.  It requires dense CSM within $6.5 \times 10^{14}$~cm of the progenitor, from a phase of advanced pre-SN mass loss beginning shortly before explosion. Spectropolarimetry of SN~2013fs shows little continuum polarization ($\sim 0.5$\%, consistent with zero), but noticeable line polarization during the plateau phase. SN~2013fr morphed from a SN~IIn at early times to a SN~II-L. After the first epoch its narrow lines probably arose from host-galaxy emission, but the
  bright, narrow H$\alpha$ emission at early times may be intrinsic to the SN.  As for SN~2013fs, this would point to a short-lived phase of strong
  CSM interaction if proven to be intrinsic, suggesting a continuum between SNe~IIn and SNe~II-L. It is a low-velocity SN~II-L like SN~2009kr, but more luminous. SN~2013fr
  also developed an infrared excess at later times, due to warm CSM dust that require a more sustained phase of strong pre-SN mass loss.
\end{abstract}


%% Paper begins here:

\begin{figure}
  \includegraphics[width=\linewidth]{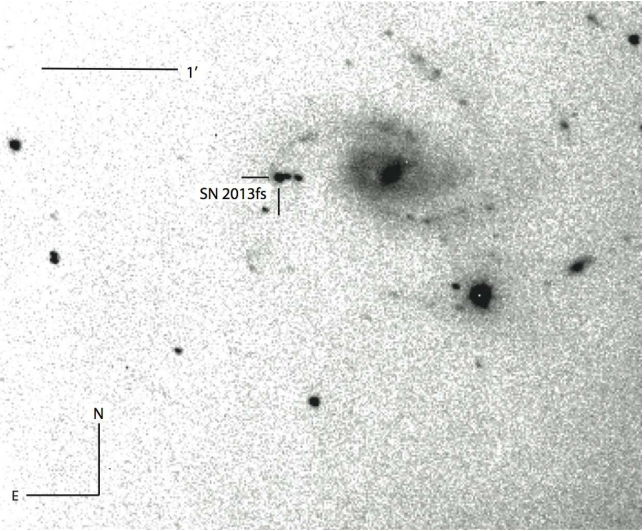}
  \caption{Unfiltered image of SN~2013fs taken by KAIT on day 1 (post-discovery; $\sim 2.5$ days post-explosion). The SN is indicated by the crosshairs, and the image scale is
    shown on the bottom left.}
  \label{fig:fsObs}
\end{figure}

%% Introduction
\section{Introduction} \label{Intro}
The expected results of core collapse in massive stars are strongly linked to the nature of the progenitor, but connections between progenitor types and supernova (SN) properties remain
uncertain. Further complicating matters, spectral and photometric features in a SN are not necessarily linked to the physics of the explosion, because they may be partly a
result of the progenitor's environment. The Type IIn SN subclass is a particularly stark example. First coined by \citet{Schlegel1990} and found by \citet{Smith2011}
to account for $\sim 9$\% of all CCSNe, SNe~IIn are characterised by strong, relatively narrow emission lines, with a general lack of broad components \citep{Filippenko1997}. These narrow
lines are not emitted by the SN ejecta; rather, they come from the ionisation of slow, dense circumstellar material (CSM) shed by the progenitor owing to mass loss during its
later evolution \citet[][and references within]{Smith2014}. They are often accompanied by broader lines, a result of shock interaction between the SN ejecta and the
dense CSM.

Owing to the fact that mass loss can result from different processes, ranging from pre-SN eruptions to binary interactions or normal radiation-driven winds, SNe~IIn are one of
the most heterogeneous types of SNe and do not result from a unique class of progenitors. The typical pre-SN mass-loss rates for progenitors that become SNe~IIn are
greater than $10^{-4}~{\rm M}_{\odot}$~yr$^{-1}$, and can in some cases be $1~{\rm M}_{\odot}$~yr$^{-1}$ or more, for superluminous SNe~IIn (SLSNe~IIn) and even some normal SNe~IIn \citep{Smith2014}. Such extreme mass loss is unlikely to be generated by the steady, radiation-driven winds associated with evolved supergiants. Among the most promising
ways to achieve such high mass-loss rates is eruptive pre-SN instability in evolved supergiant stars, known to occur in luminous blue variables (LBVs) such as $\eta$ Carinae (with a very high total
ejecta mass $> 10~{\rm M}_{\odot}$; \citealt{Smith2003}) and P Cygni (with a more modest ejecta mass of $\sim 0.1~{\rm M}_{\odot}$; \citealt{SmithHart2006}). Eruptive instability has been
proposed to explain SNe~IIn before \citep{Smith2006}, and an erupting LBV becoming a SN~IIn has likely been observed in at least one case (e.g., the 2012 outburst of
SN~2009ip; \citealt{Mauerhan2013}); massive stars becoming SNe~Ibn shortly after leaving the LBV phase have been observed as well \citep{Foley2007}.

While SNe~IIn represent the most extreme pre-SN mass loss, some normal SNe also show signs of strong winds or eruptions. Those classified as SNe~IIn throughout their entire
evolution have a total CSM in the range 0.1--1~M$_{\odot}$ for normal SNe~IIn, and as much as 10--20~M$_{\odot}$ or more for SLSNe~IIn \citep{Smith2014}.
\citet{Smith2015} argued that the degree of mass loss required to generate the characteristic features of SNe~IIn extends to lower mass-loss rates, and that some SNe~II would be classified as SNe~IIn {\it only if discovered sufficiently early after explosion}, evolving into other types (SNe~II-P, SNe~II-L) as they age.

\begin{figure}
  \includegraphics[width=\linewidth]{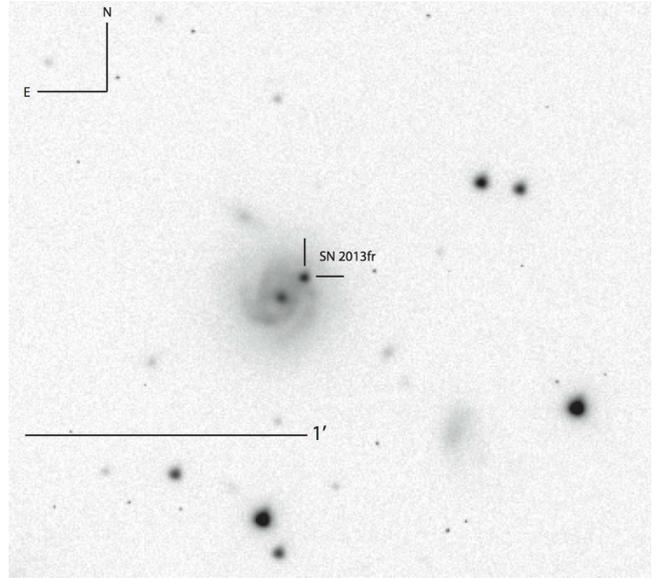}
  \caption{{\it R}-band Lick Nickel telescope image of SN~2013fr, obtained 10 days post-discovery. The SN is indicated by the crosshairs,
    and the image scale is shown on the bottom left.}
  \label{fig:frObs}
\end{figure}

Some young SNe display transient Wolf Rayet (WR)-like high-ionisation emission lines, present only if discovered {\it very} early (less than a week post-explosion). The
earliest such example was reported for SN~1983K by \citet{Niemala1985}, and this was followed a decade afterward by \citet{Benetti1994}, who observed the same lines in
early-time spectra of SN~1993J. The cause of these lines has been interpreted in different ways by different authors. Early papers regarded the He\,\textsc{ii} $\lambda$4686 in
particular as likely the result of highly blueshifted H$\beta$ emission. Later, \citet{Quimby2007} describe them as the result of X-rays coming from the shock breakout (SBO)
and subsequent shock interaction between the SN ejecta and CSM. Most recently, \citet{GalYam2014} and \citet{Khazov2016} ascribe them to the post-SBO ultraviolet (UV) flash.

\citet{Leonard2000} reported both early narrow lines and a high degree of continuum polarization in SN~1998S, interpreting the polarization as evidence for asphericity;
\citet{Shivvers2015} later published an even earlier spectrum of SN~1998S, finding the WR-like lines to be stronger. These lines have been observed in several SN types:
\citet{Quimby2007} found them in SN~2006bp (II-P), and later \citet{GalYam2014} and \citet{Smith2015} identified these features in SN~2013cu (IIb) and PTF11iqb (IIn). \citet{GalYam2014} proposed, based on the similarity of the spectral features to those of a WN6h spectral type, that SN~2013cu resulted from the explosion of a star having some characteristics of a WR star.
However, subsequent work \citep{Groh2014,Smith2015} ruled out a WR progenitor, finding that an LBV or yellow hypergiant was more likely. In the last year,
\citet{Khazov2016} concluded that early, rapidly fading narrow spectral lines could appear in up to 18\% of young SNe. Most of these objects are classified as SNe~IIb
or SNe~IIn, with few known SNe~II-P showing such lines.

The existence of any intrinsic distinction between SNe~II-P and II-L has been a subject of debate \citep{Filippenko1997,Arcavi2012,Sanders2015,Valenti2016,Anderson2014}. Spectroscopically, SNe~II-L often show weaker P Cygni and Ca\,\textsc{ii} absorption than SNe~II-P, which may be evidence of CSM
interaction in the former due to reionisation of the outer SN ejecta \citep{Smith2015}. Adding to the observational evidence, recent hydrodynamical models by
\citet{Morozova2016} find that the light curves of SNe~II-P and II-L can be fit well by RSGs with dense CSM.

In this paper we present optical photometry and spectroscopy of SN~2013fs (shown in Figure \ref{fig:fsObs}) and
SN~2013fr (shown in Figure \ref{fig:frObs}). Both of these objects showed narrow H$\alpha$ emission at early times, and were initially classified as SNe~IIn, but the narrow component caused by CSM interaction
later weakened or even disappeared.

SN~2013fs (additional designations: iPTF13dqy and PSN J23194467+1011045) was discovered at $R = 16.5$~mag by the Itagaki Astronomical Observatory on 2013 Oct. 7.468 (UT dates are used throughout this paper) in
an outer spiral arm of NGC 7610 \citep{CBET3671}. Based on the host-galaxy redshift of $z = 0.01178$ (3554~km~s$^{-1}$; NED\footnote{NED: The NASA/IPAC Extragalactic Database (NED)
  is operated by the Jet Propulsion Laboratory, California Institute of Technology, under contract with the National Aeronautics and Space Administration (NASA).}) and five-year WMAP parameters \citet{Komatsu2009}, we adopt a
distance of $50.6 \pm 0.9$~Mpc ($m - M = 33.52 \pm 0.04$~mag) with $E(B-V) = 0.0347$~mag \citep{Schlafly2011}. SN~2013fs is located 49{\arcsec} east and 2{\arcsec} south of the host nucleus
(separation $\sim 12$~kpc), and was initially classified as a probable SN~IIn based on a spectrum obtained on 2013 Oct. 8 (day 1). SN~2013fs
was reclassified on 2013 Oct. 24 (day 17) with additional WiFeS data as a SN~II-P, with SNID \citep{Blondin07} providing a best match to SN~1999em \citep{ATEL}.

SN~2013fr was discovered at $I = 16.2$~mag by the Catalina Sky Survey on 2013 Sep. 28.42 at a projected separation of $\sim 6.7${\arcsec} ($\sim 2.5$--3~kpc)
from the nucleus of MCG+04-10-24, in an inner spiral arm. Initially designated as PSN J04080235+2317394, SN~2013fr was classified as a SN~IIn using spectra from the F. L.
Whipple Observatory 1.5~m telescope obtained on day 1 \citep{CBET3666}. Based on the host $z = 0.020941$ (6278~km~s$^{-1}$) from NED and
five-year WMAP parameters \citet{Komatsu2009}, we adopt a distance of
$87.0 \pm 1.6$~Mpc ($m - M = 34.70 \pm 0.04$~mag) with $E(B-V) = 0.2308$~mag \citep{Schlafly2011}. This yields a discovery {\it I}-band absolute magnitude of $-18.48$.

%% Observations & Data
\section{Observations and Data} \label{obs}

\subsection{SN~2013fs\protect\footnote{Some facilities obtained data for both SNe. In these cases, the data obtained with those facilities are mentioned in this section
only.}}\label{obs:13fs}

\subsubsection{Lick KAIT/Nickel photometry} \label{obs:KAIT}
After each discovery, both SNe were observed by the Katzman Automatic Imaging Telescope \citep[KAIT;][]{filippenko2001} at Lick Observatory.
KAIT obtained photometry of SN~2013fs in the clear filter for 61 epochs from 0 to 109 days post-discovery, and 26 epochs of {\it BVRI}
photometry from 3 to 49 days post-discovery. No source was detected at the location of SN~2013fs on 2013 Oct. 05.27 (two days pre-discovery)
down to a 3$\sigma$ unfiltered limit $\sim 19.24$~mag. The resulting data are listed in Table \ref{tab:lickfs} for the filtered photometry
and Table \ref{tab:clearfs} for the unfiltered photometry.

KAIT photometry of SN~2013fr covers 9 epochs from day 22 to day 58 in the {\it BVRI} filters, and the Nickel telescope at Lick observed
SN~2013fr for 8 epochs in the same filters; these data are listed together in Table \ref{tab:lickfr}. All photometry is corrected for Milky
Way line-of-sight extinction $E(B-V) = 0.0347$~mag \citep{Schlafly2011}.

The data was reduced with the KAIT standard photometry pipeline \citet{Ganeshalingam2010}, using template subtraction to remove host galaxy
contamination, aperture photometry using Landolt standards is used to calibrate to the standard photometry system. All KAIT/Nickel data are
reported on the Vega system.

\subsubsection{Kuiper photometry} \label{obs:Kuiper}
Four epochs of late-time {\it BVRI} photometry of SN~2013fs were obtained by the AZTEC (Arizona Transient Exploration and Characterization) collaboration with the Kuiper 61~inch
telescope on Mt. Bigelow, using the Mont4k CCD at the {\it f}/13.5 Cassegrain focus, and a $9.73' \times 9.73'$ field of view with 0.42{\arcsec} pixel$^{-1}$.The seeing
ranged between 1{\arcsec} and 2{\arcsec}. No source was detected at the location of the SN; limiting magnitudes were obtained using the method described by \citet{Bil2015}.
The late-time data are given in Table \ref{tab:kuiper}.

\subsubsection{{\it Swift} UVOT photometry} \label{obs:uvot}
After discovery, SN~2013fs was added to the queue for the UltraViolet and Optical Telescope (UVOT) aboard {\it Swift} \citep{Rom2005}, and observed for 21 epochs with all
filters. The base UVOT data were taken from the {\it Swift} Optical/Ultraviolet Supernova Archive \citep{Brown2014}. We obtained a template image of NGC 7610 on 2016
July 14, and the data were reduced using the method of \citet{Brown2009}, subtracting host-galaxy count rates and using the revised UV zeropoints and time-dependent
sensitivity from \citet{Bre2011}. The resulting photometry is shown in Table \ref{tab:uvot}.

\begin{figure*}
  \centering
  \includegraphics[width=\textwidth]{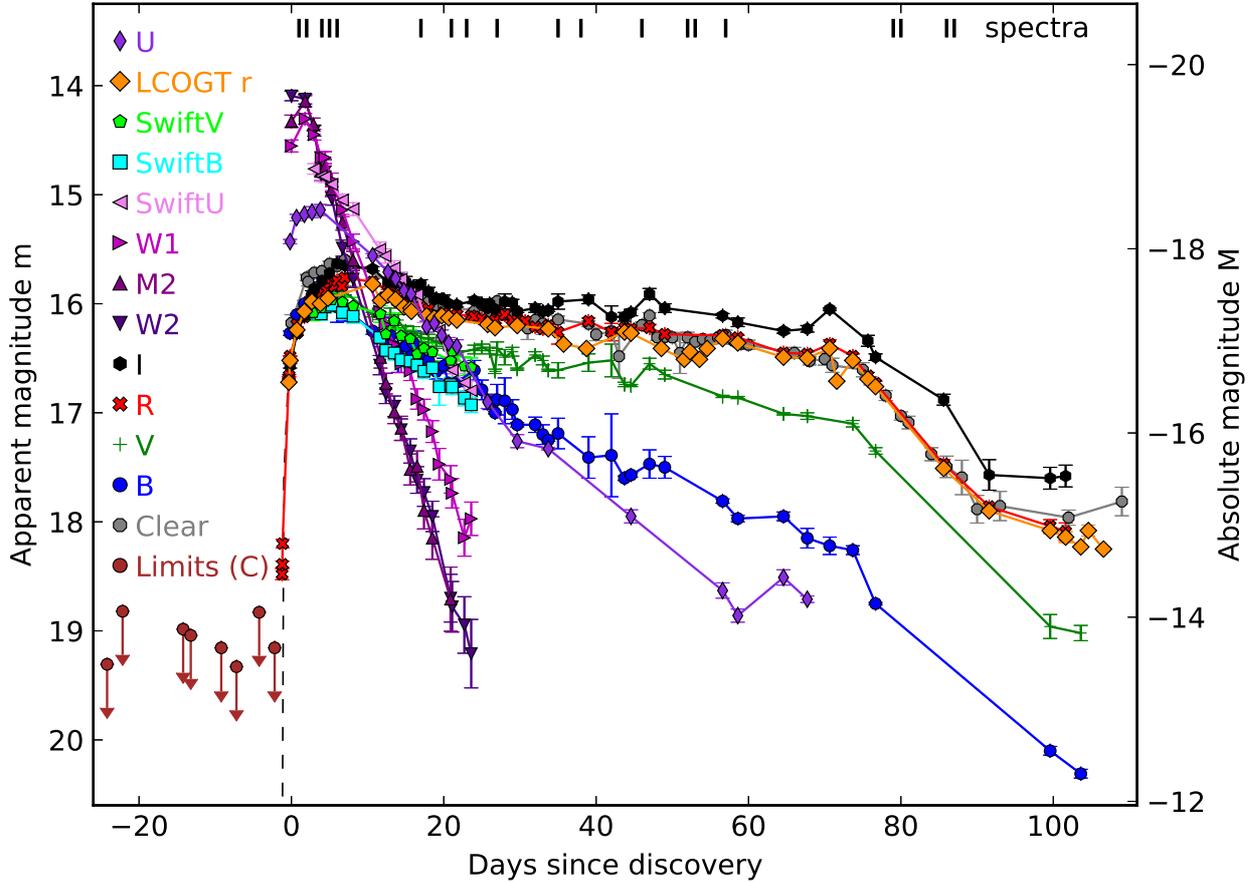}
  \caption{All early-time UV and optical light curves of SN~2013fs. The unfiltered data in grey are roughly the photometric equivalent of an {\it R} filter. All photometry have been corrected for Milky Way reddening with $E(B-V) = 0.0347$~mag, from \citet{Schlafly2011}. Magnitudes
  are reported on the Vega system.}
  \label{fig:fsphot}
\end{figure*}

\begin{figure}
  \includegraphics[width=\columnwidth]{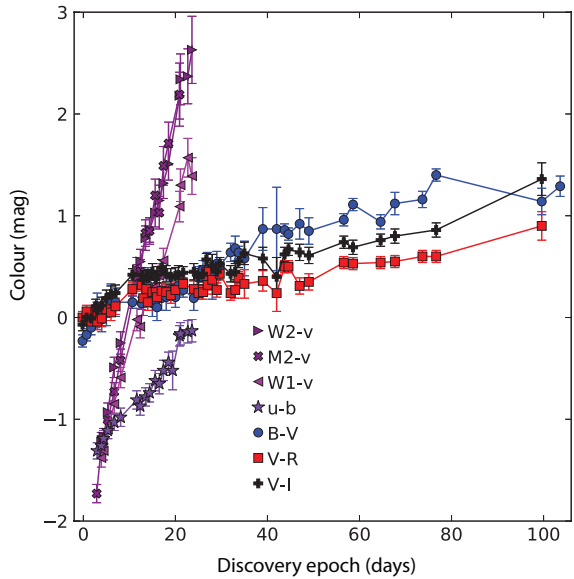}
  \caption{Colour evolution of SN~2013fs. The {\it B--V}, {\it V--R}, and {\it V--I} evolution is largely flat with time. All photometry has
  been corrected for Milky Way reddening with $E(B-V) = 0.0347$~mag.}
  \label{fig:fscolors}
\end{figure}

\begin{figure*}
  \centering
  \includegraphics[width=0.8\textwidth]{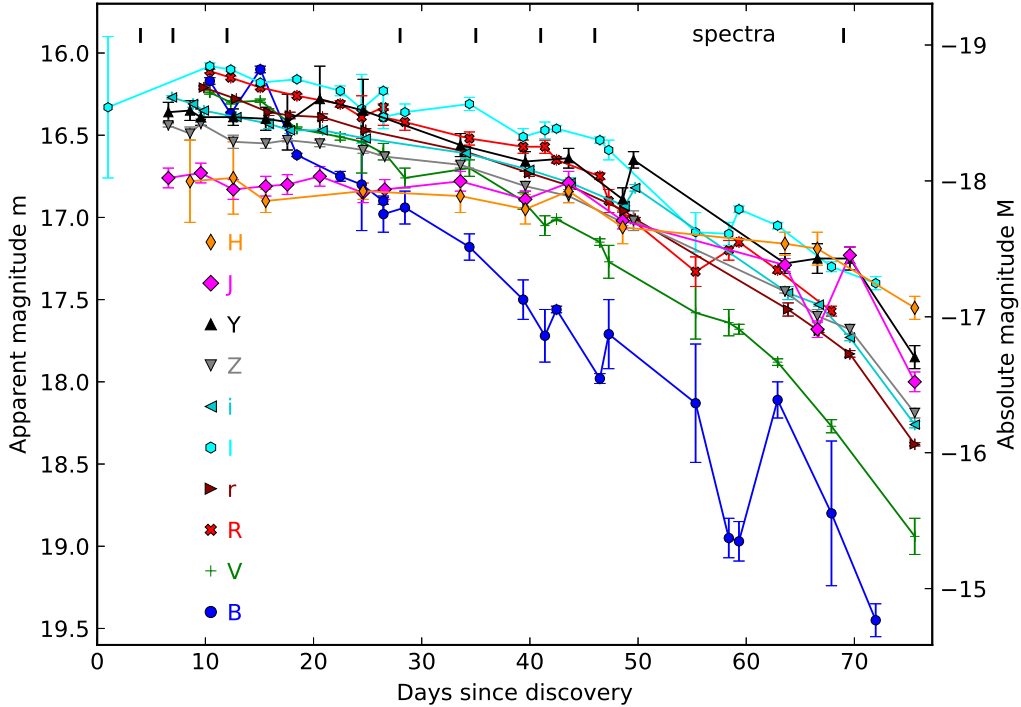}
  \caption{Optical and near-infrared light curves of SN~2013fr. The BVRI photometry is on the Vega system, and the rizYJH have been converted
    to the AB system. All points in this plot have been corrected for Milky Way reddening with $E(B-V) = 0.2076$~mag, from
    \citet{Schlafly2011}.}
  \label{fig:frphot}
\end{figure*}

\begin{figure*}
  \includegraphics[width=\linewidth]{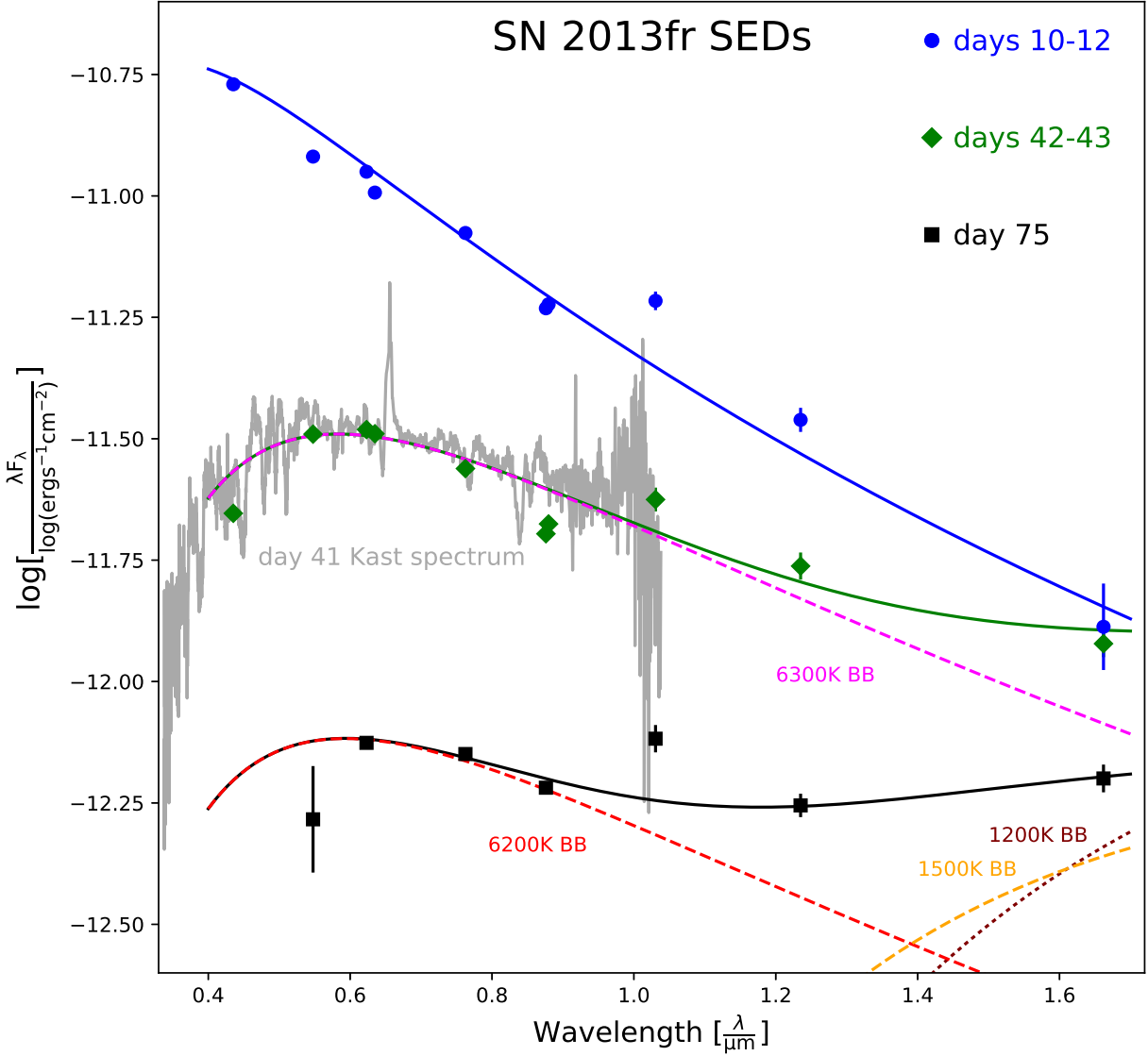}
  \caption{SEDs of SN~2013fr for days 10--12 (blue), days 42/43 (green), and day 75 (black) post-discovery. Each SED has an associated blackbody function having the same
    colour: single temperature for the first SED, and double temperature for the others. The $Y$-band photometry at 1 $\mu$m is calibrated to the UKIDDS $Y$ band, which may
    have caused a systemic shift in brightness. All points are calibrated for Milky Way reddening ($E(B-V) = 0.2076$~mag).}
  \label{fig:frSED}
\end{figure*}

%***Chris: In Figure 7, the ordinate units should be 
% ``(erg s^-1 cm^-2 Ang^-1)'' (not italicized; use Ang symbol).
%  The abscissa units should be simply ``(Ang)'' (symbol, and
% not italicized; don't give the wavelength range because it looks
% like part of the unit, and people can see the wavelength range
% in the figure.
%   There are a number of strong, narrow emission lines that
% are NOT labeled. What are they?! I'm talking about two
% lines near 5150 Ang and three lines near 6000 Ang. I suggest
% you label them and mention them in the text.

\begin{figure*}
  \centering
  \includegraphics[width=\textwidth]{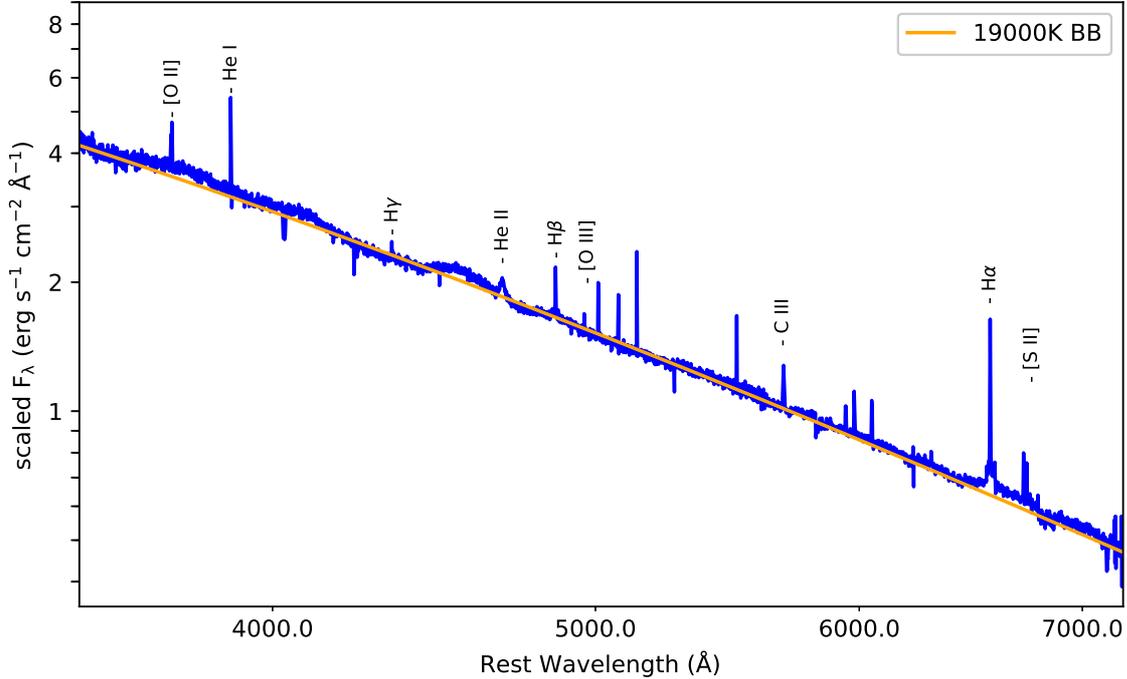}
  \caption{Day 1 (post-discovery) spectrum of SN~2013fs overplotted with a 19,000~K blackblody, with spectral features of interest marked.
    The spectrum has been dereddened and redshift corrected. We suspect that the unmarked narrow features from $\sim$5000--6000\AA are
    telluric sky lines that were incompletely subtracted; they are all narrow and unresolved, and they coincide with wavelengths of expected
  sky lines. Further, they do not match the wavelengths of narrow emission lines seen in other SNe II~P discovered early.}
  \label{fig:day1}
\end{figure*}

%***Chris: In Figure 8, the ordinate units should be 
% ``(erg s^-1 cm^-2 Ang^-1)'' (not italicized; use Ang symbol).
% And I would write ``Scaled F_\lambda'' instead of
% putting ``scaled'' in parentheses.

%***Chris: In Figure 9, the ordinate units should be 
% ``(erg s^-1 cm^-2 Ang^-1)'' (not italicized).
% And I would write ``Scaled F_\lambda'' instead of
% putting ``scaled'' in parentheses.

\begin{figure*}
  \centering
  \begin{minipage}{0.45\textwidth}
    \includegraphics[width=1.15\columnwidth, height=0.65\textheight]{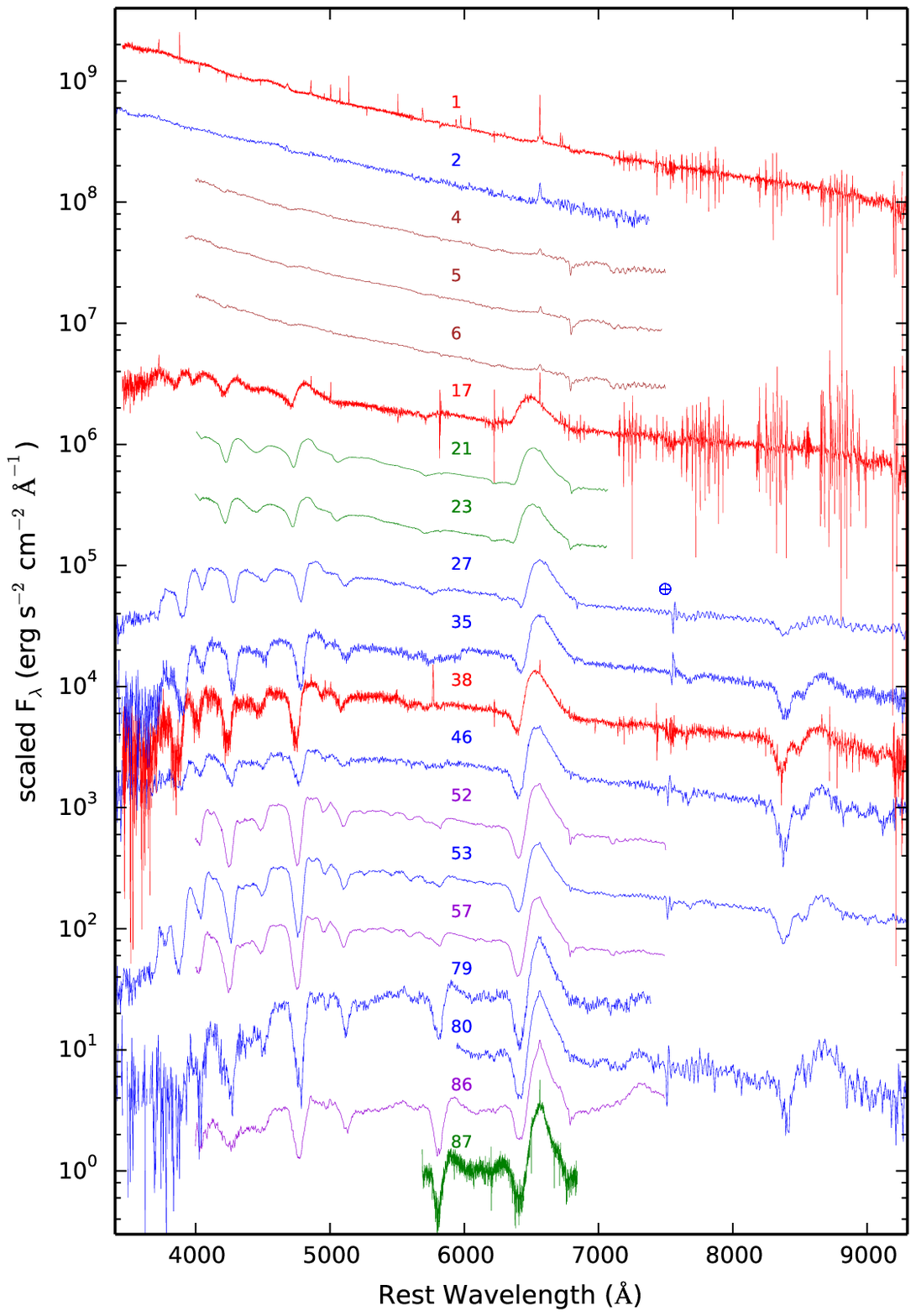}
    \caption{Spectra of SN~2013fs, corrected for Milky Way reddening, from WiFeS (red), PESSTO (blue), Kuiper (brown), Bok (purple), and the MMT (green). $\oplus$ markings indicate uncorrected telluric absorption.}
    \label{fig:fsallspec}
  \end{minipage}%
  \hfill
  \begin{minipage}{0.45\textwidth}
    \includegraphics[width=1.15\columnwidth, height=0.65\textheight]{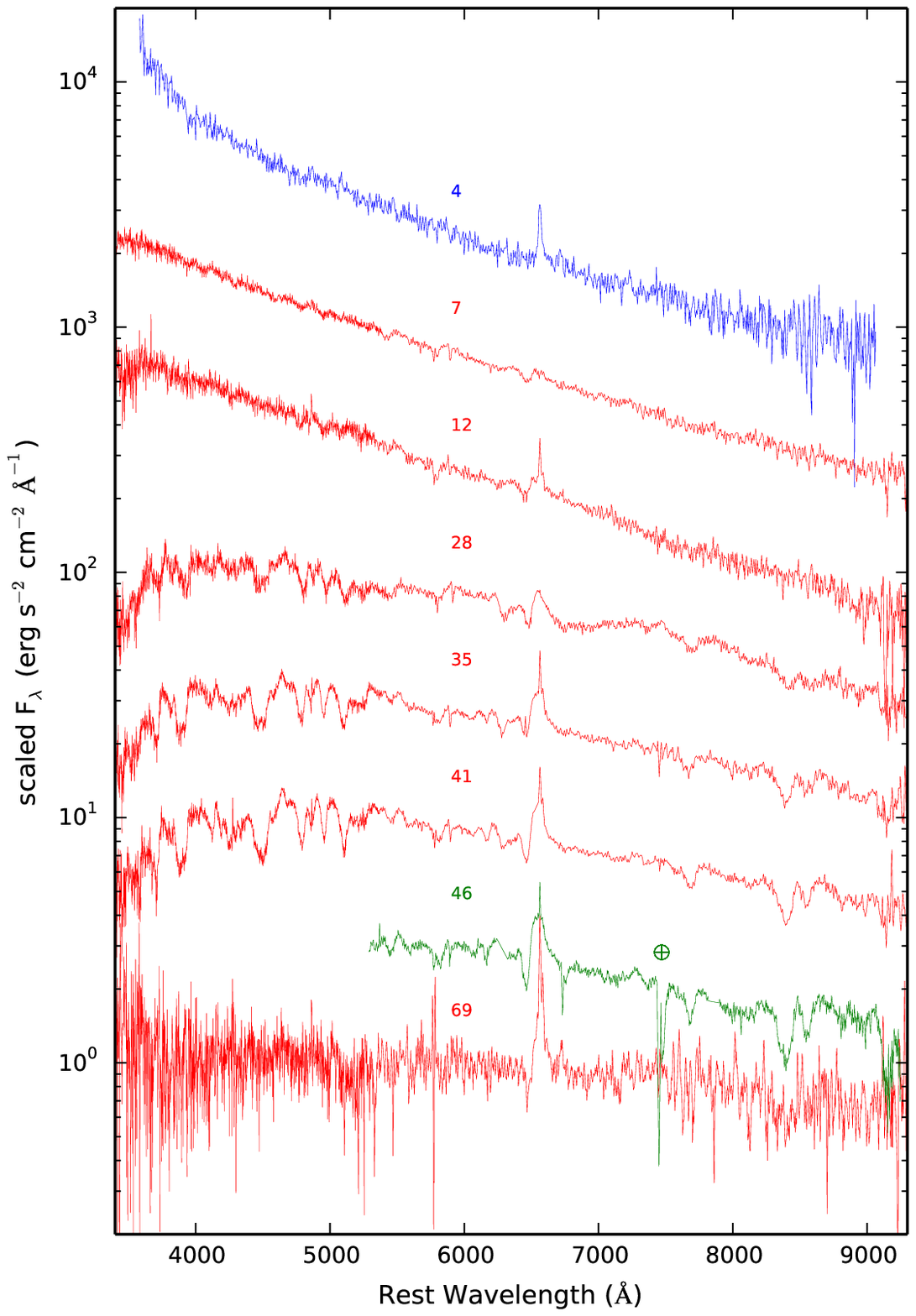}
    \caption{All spectra of SN~2013fr, corrected for Milky Way reddening. Red spectra are from the Lick Shane telescope, green from Magellan, and blue from PESSTO;
      the $\oplus$ marks indicate uncorrected telluric absorption.}
    \label{fig:frallspec}
  \end{minipage}
\end{figure*}

\begin{figure*}
  \centering
  \includegraphics[width=\textwidth]{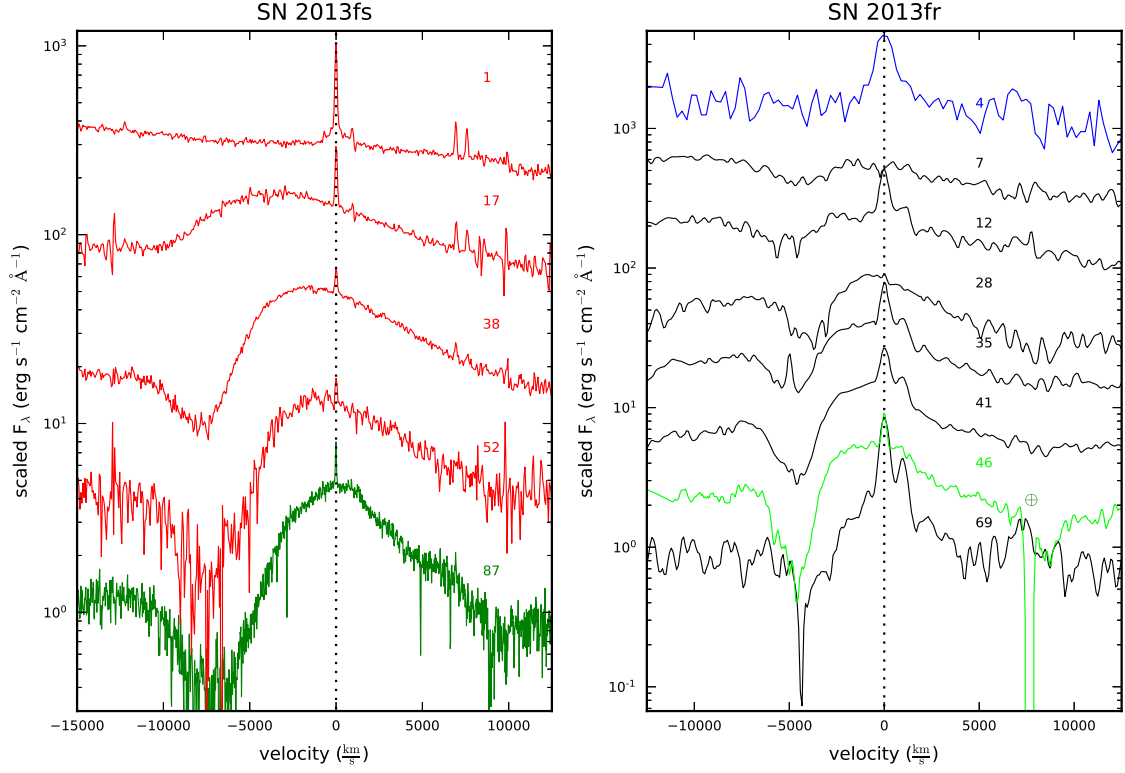}
  \caption{H$\alpha$ line profile of both SNe; SN~2013fs is on the left, SN~2013fr on the right. Telluric features are marked with $\oplus$. Spectra are from WiFeS (red),
    MMT (dark green), PESSTO (blue), Lick/Shane (black), and Magellan (light green).}
  \label{fig:halpha}
\end{figure*}

\begin{figure*}
  \centering
  \includegraphics[width=\textwidth, height=0.35\textheight, keepaspectratio]{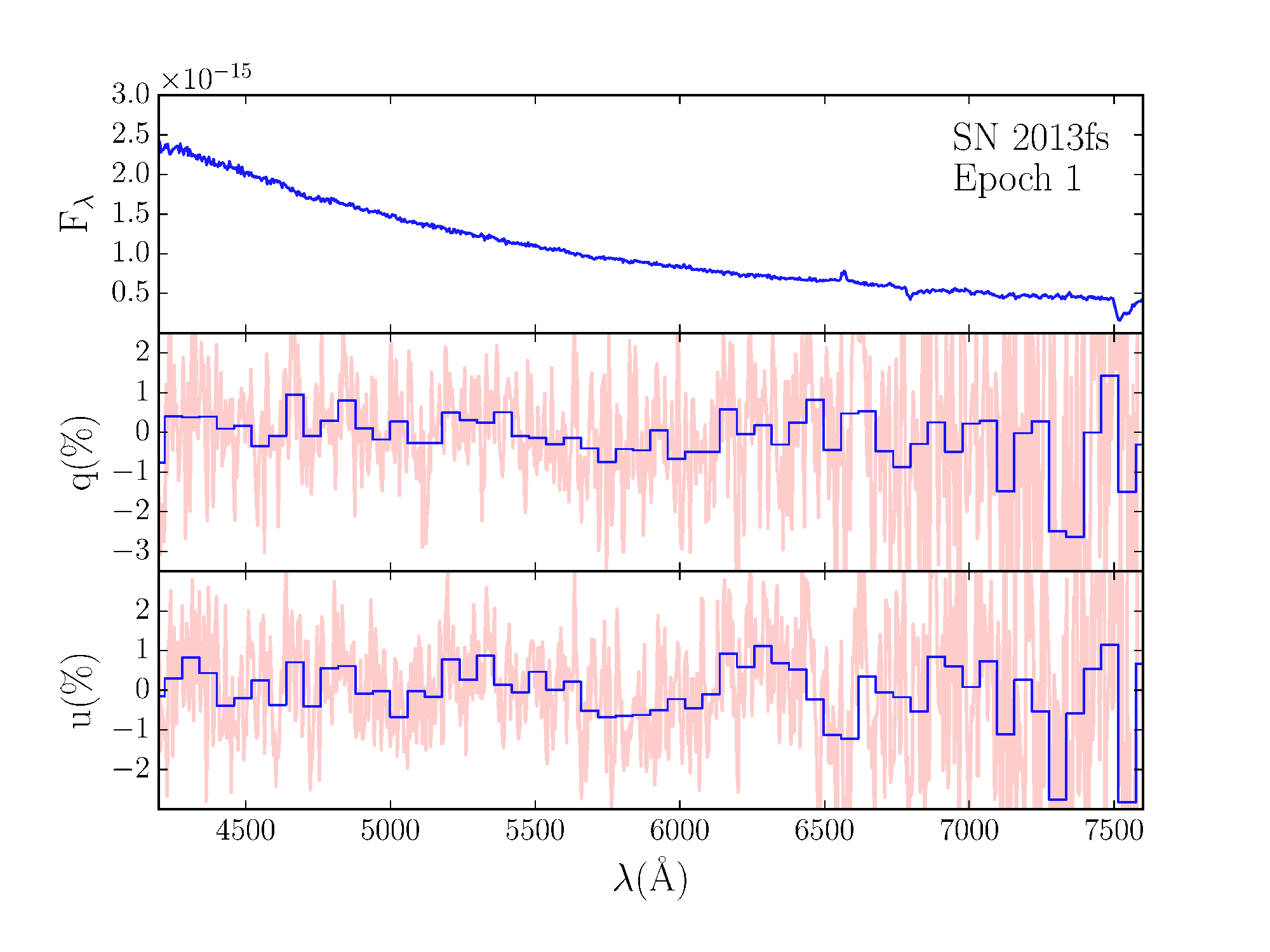}
  \caption{Kuiper spectropolarimetry of SN~2013fs, days 4--6 post-discovery.}
  \label{fig:specpol1}
\end{figure*}

\begin{figure*}
  \centering
  \includegraphics[width=\textwidth, height=0.35\textheight, keepaspectratio]{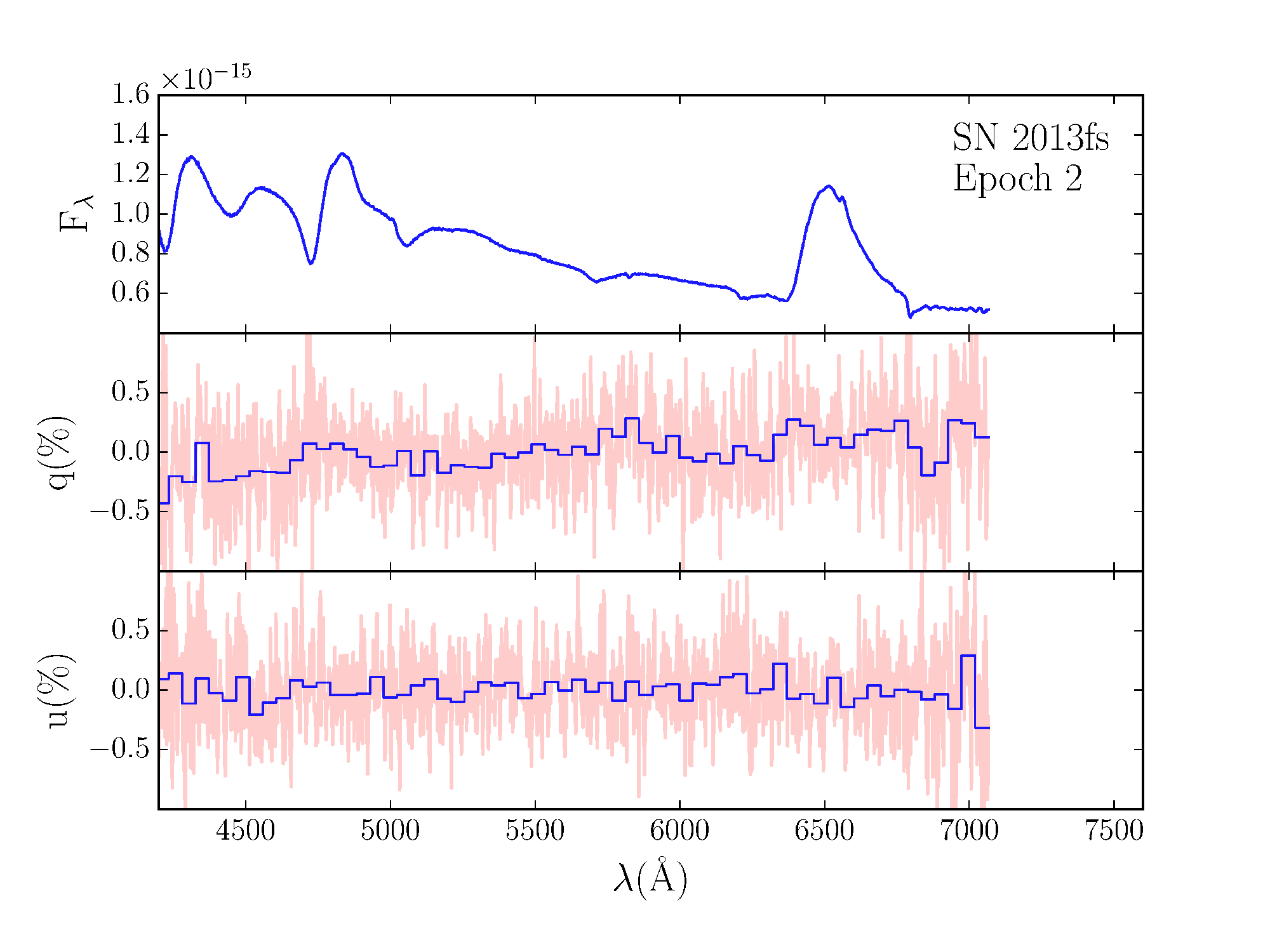}
  \caption{MMT spectropolarimetry of SN~2013fs, days 21 and 21 post-discovery.}
  \label{fig:specpol2}
\end{figure*}

\begin{figure*}
  \centering
  \includegraphics[width=\textwidth, height=0.35\textheight, keepaspectratio]{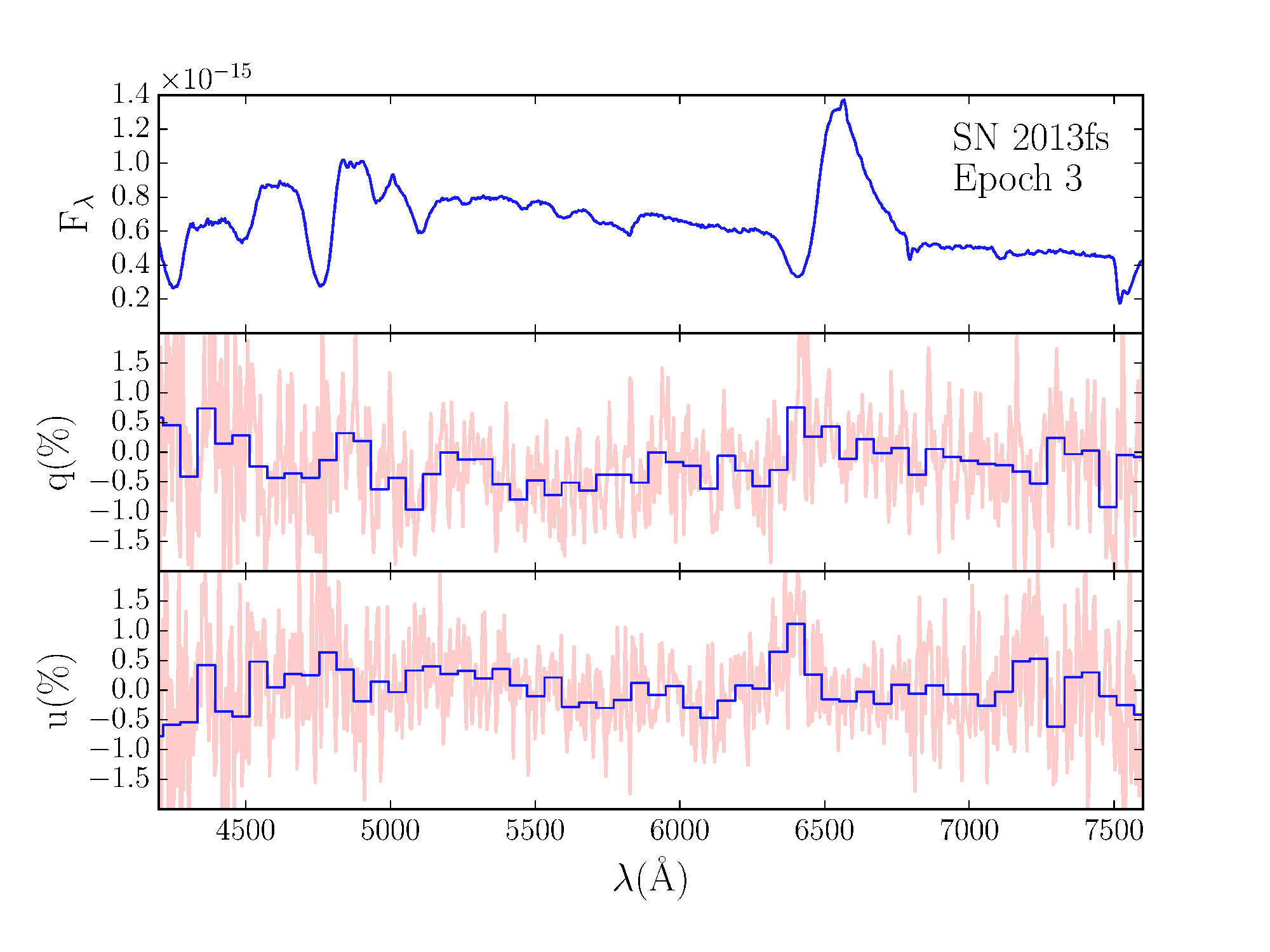}
  \caption{Bok spectropolarimetry of SN~2013fs, days 52 and 57 post-discovery.}
  \label{fig:specpol3}
\end{figure*}

\begin{figure*}
  \centering
  \includegraphics[width=\textwidth, height=0.35\textheight, keepaspectratio]{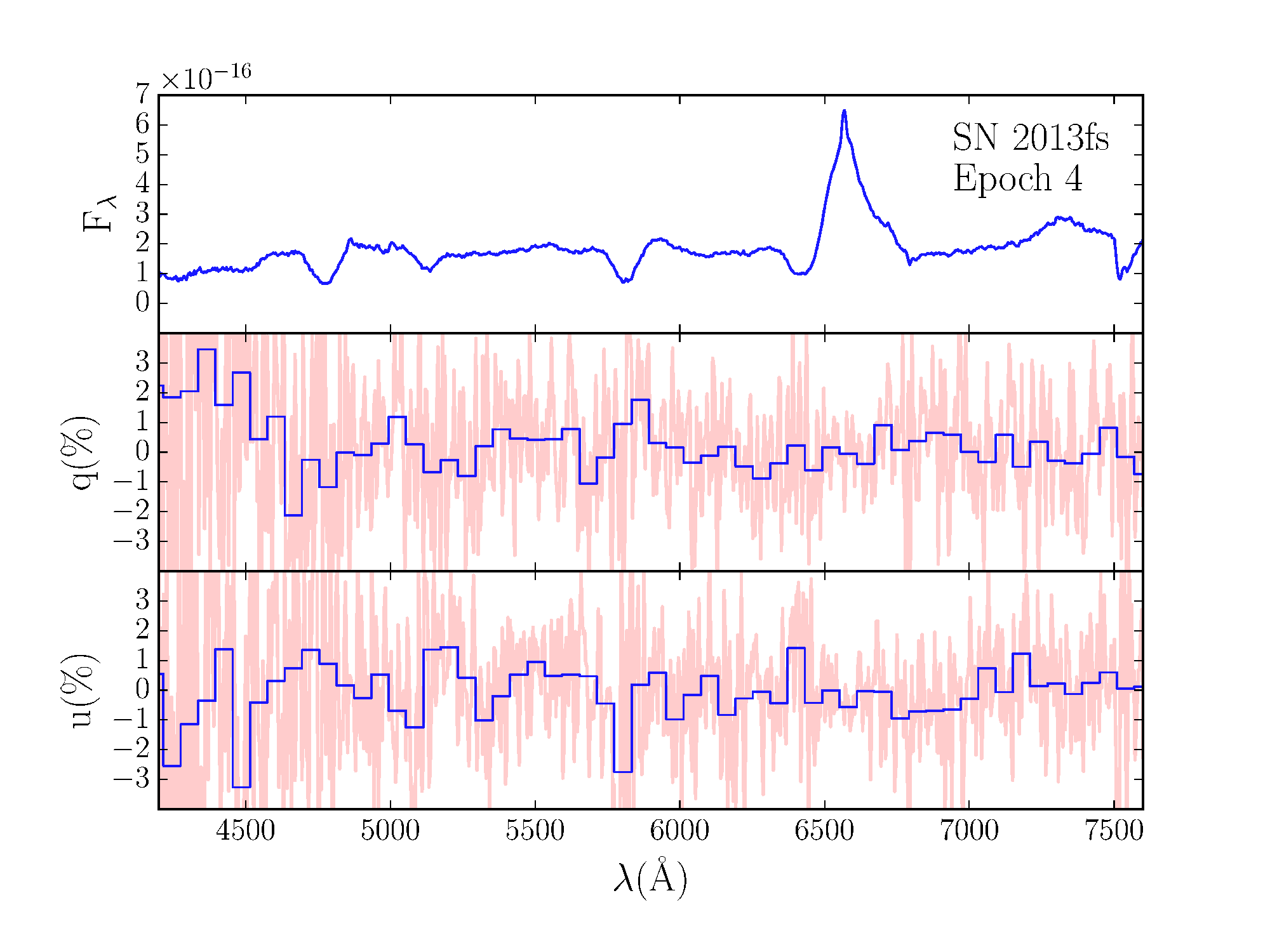}
  \caption{Bok spectropolarimetry of SN~2013fs, days 86 and 87 post-discovery.}
  \label{fig:specpol4}
\end{figure*}

\begin{figure*}
  \includegraphics[width=0.56\textwidth]{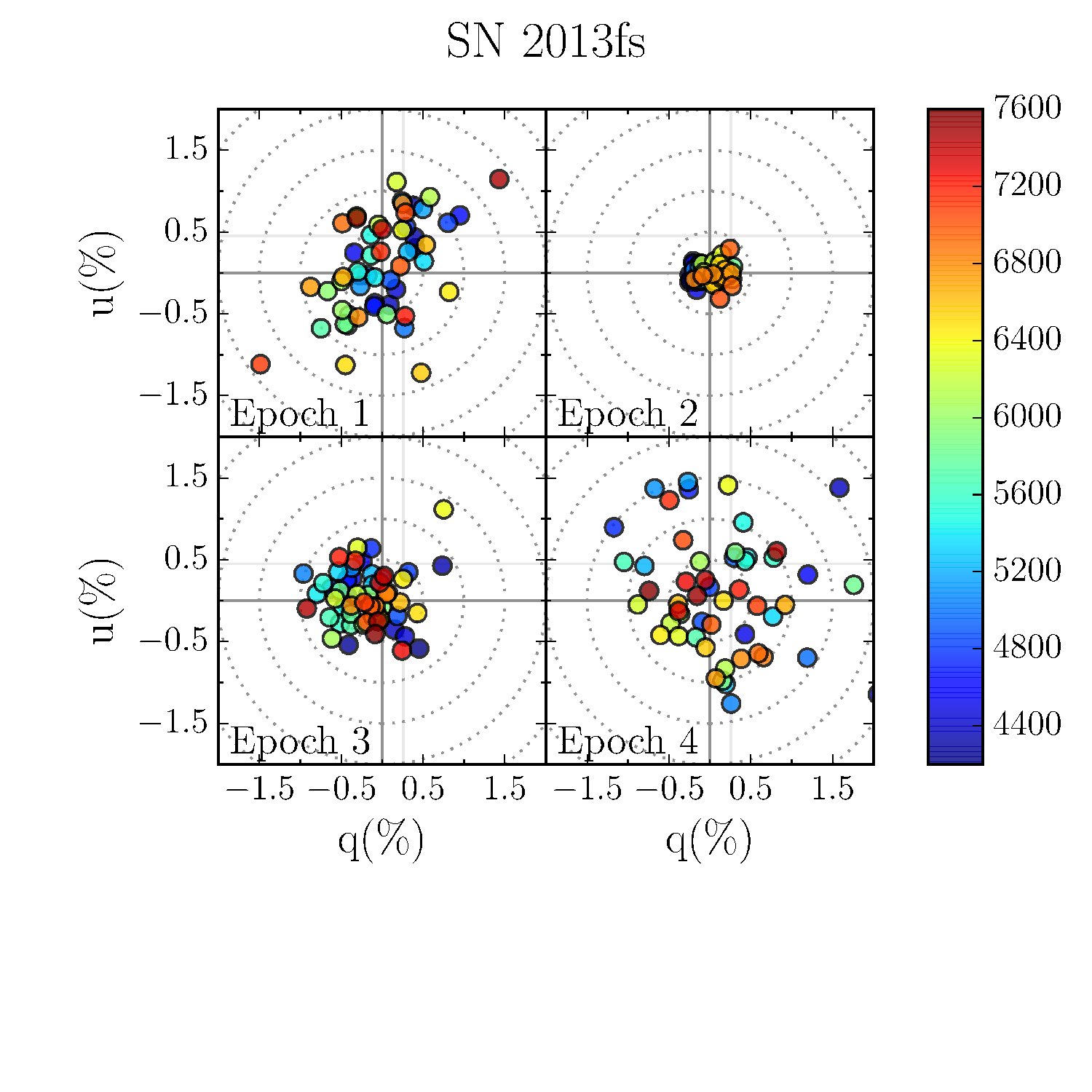}
  \caption{Line polarization in SN~2013fs plotted in the ($q,u$) plane. Points are colour-coded by wavelength (\AA) as described in Section \ref{res:specpol}.}
  \label{fig:quplane}
\end{figure*}

\begin{figure}
  \centering
  \includegraphics[width=\columnwidth]{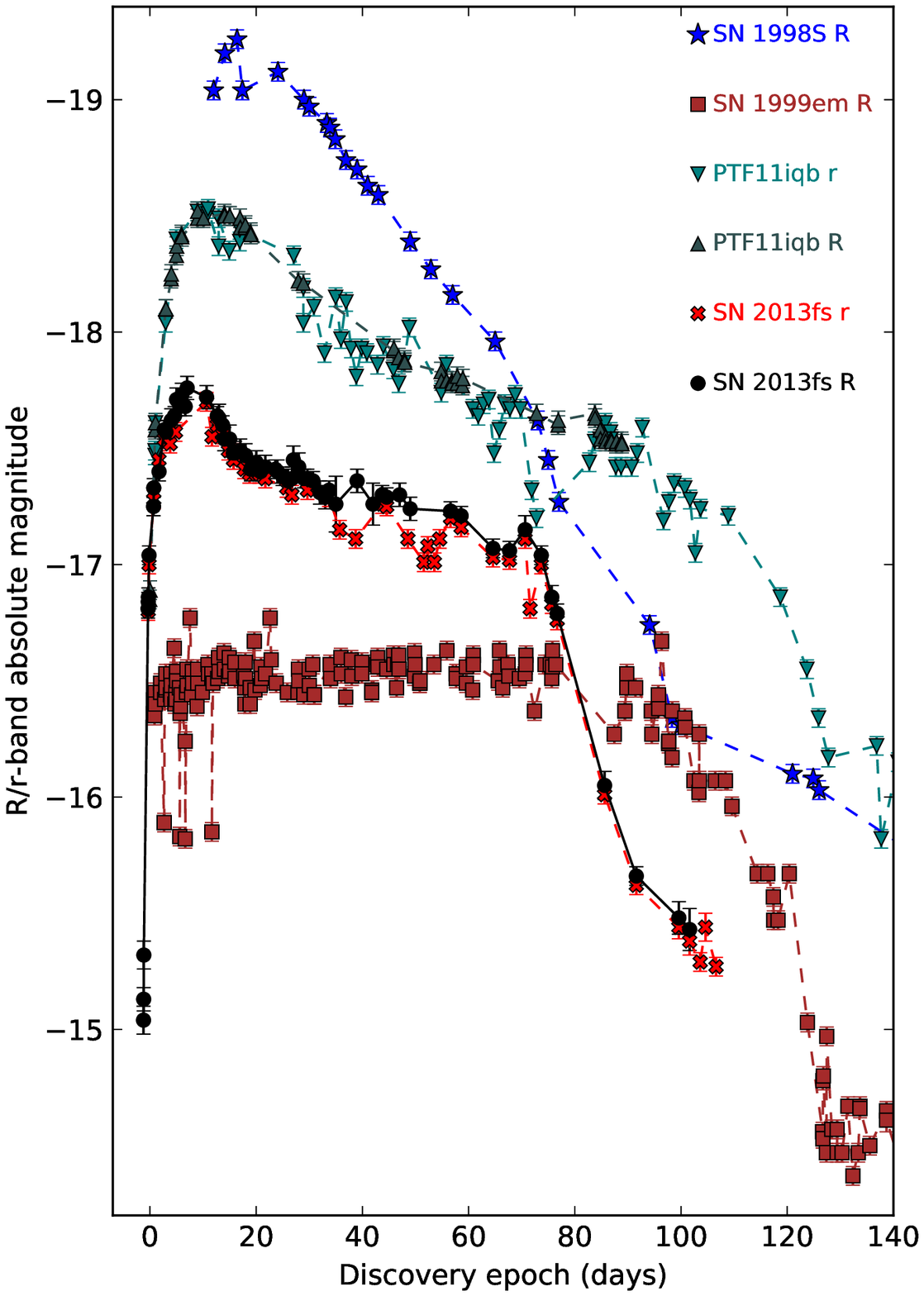}
  \caption{$r/R$-band comparison of SN~2013fs with other SNe discussed in the literature. All photometry of SN~1999em (brown; \citealt{Leonard2002}; \citealt{Faran2014-2};
    \citealt{Galbany2016}), SN~1998S (blue;~\citealt{Fassia2001}), and PTF11iqb (teal and dark for the {\it r} \& {\it R} bands, respectively; \citealt{Smith2015}) were
    retrieved from the Open Supernova Catalog. All photometry is corrected for Milky Way reddening. Days are plotted with respect to discovery, but all of these SNe were discovered within a few days of explosion.}
  \label{fig:comparefs}
\end{figure}

\begin{figure*}
  \includegraphics[width=\textwidth]{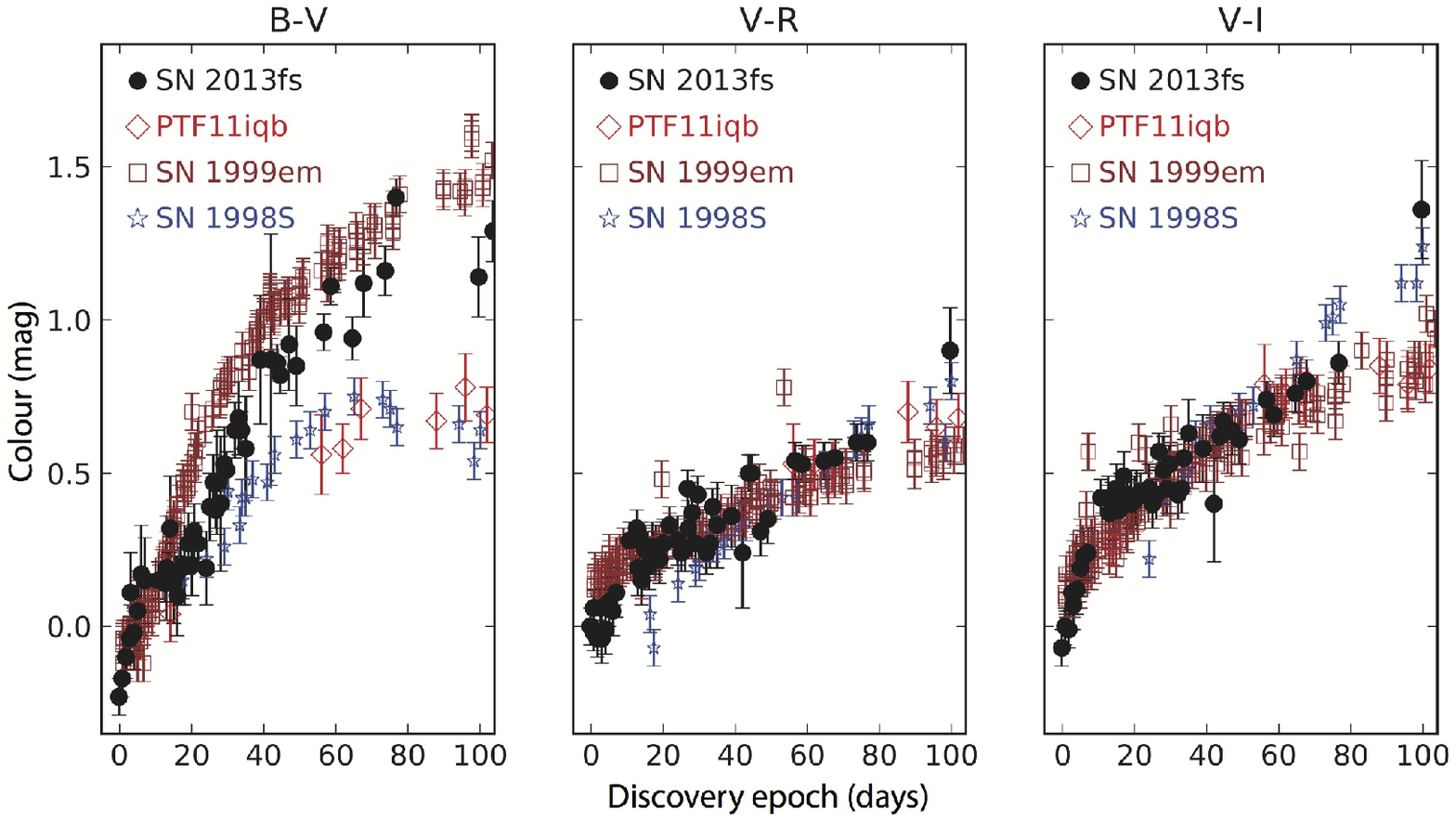}
  \caption{Comparative colour evolution of SN~2013fs (black), SN~1998S (blue), and SN~1999em (brown); each has been corrected for MW reddening with $E(B-V) = 0.0347$~mag \citep{Schlafly2011}.}
  \label{fig:fscolorcompare}
\end{figure*}

\subsubsection{Spectroscopy} \label{obs:fsspec}
Eight low-to-moderate resolution spectra and one higher-resolution spectrum covering the first 87 days post-explosion were obtained of
SN~2013fs, and they are supplemented with additional archival spectra (see Section \ref{obs:archspec}). Three spectra were obtained during
the first week using the SPol spectropolarimeter on the Kuiper telescope, two on days 21 and 23 using SPol at MMT Observatory, and three more
between days 52 and 86 with SPol on the 2.1~m Bok telescope. One high-resolution spectrum was taken on day 87 using the Bluechannel
spectrograph at the MMT. The technical details of these observations appear in Table \ref{tab:spectra}.

All spectra were reduced using standard reduction techniques as described by \citet{Foley2003} and references therein. Wavelength calibration
was done at the 2.1~m Bok, 61" Kuiper, and 6.5~m MMT telescopes using a HeNeAr lamp, and flux calibration using spectra of standard stars
taken on the night of the observations.

\subsubsection{Spectropolarimetry} \label{obs:specpol}
As part of the Supernova Spectropolarimetry (SNSPOL) Project\footnote{\url{http://grb.mmto.arizona.edu/~ggwilli/snspol/}}, we observed SN~2013fs during four epochs using
the CCDImaging/Spectropolarimeter \citep{Schmidt1992} mounted on either the 61~inch Kuiper, 6.5~m MMT, or 90-inch Bok telescopes. These correspond to the days 4--6, day 21/23, and days 52, 57,
and 86 spectra in the top half of Table \ref{tab:spectra}, respectively. At the 90~inch and 61~inch telescopes, we used a $3.0\arcsec$ slit with the 600 lines mm$^{-1}$ grating blazed at $11.35^{\circ}$
(5819~{\AA}). This configuration provides a spectral coverage of 4970~{\AA} at a full-width at half-maximum intensity (FWHM) resolution of 19.5~{\AA}. At the MMT we used either a $1.9\arcsec$ slit or a $2.8\arcsec$
slit with the 964 lines mm$^{-1}$ grating.  This configuration provides a spectral coverage of 3140~{\AA} with a resolution of 19.8~{\AA} ($1.9\arcsec$ slit) or 29.1~{\AA} ($2.8\arcsec$
slit).

Effects of pixel-to-pixel variations are reduced by obtaining a pair of exposures with the ordinary and extraordinary traces swapped on the detector for both linear Stokes
parameters, $(Q_{1},Q_{2})$ and $(U_{1}, U_{2})$. During data reduction, those exposures are combined into a single $Q$ image and a single $U$ image. To reduce variations caused by
waveplate orientation, each exposure is an integration over four equivalent waveplate rotation angles as follows:
\begin{equation}
\begin{split}
Q_1:& \thickspace 67.5^{\circ}, 157.5^{\circ}, 247.5^{\circ}, 337.5^{\circ}\\
Q_2:& \thickspace 22.5^{\circ}, 112.5^{\circ}, 202.5^{\circ}, 292.5^{\circ}\\
U_1:& \thickspace 45^{\circ}, 135^{\circ}, 225^{\circ}, 315^{\circ}\\
U_2:& \thickspace 0^{\circ}, 90^{\circ}, 180^{\circ}, 270^{\circ}
\end{split}
\end{equation}

During each observing run we obtained a standard set of calibration data which include He-Ne-Ar comparison-lamp spectra and flatfield spectra for every slit. To measure and
correct for the polarimetric efficiency of the instrument, we also obtained a full $Q, U$ sequence of a continuum source through a nicol prism. Observations of unpolarized
flux standards (BD+28~4211 and G191-B2B) and polarized standards (Hiltner~960, VI~Cyg~$\#12$, BD+59~389, and HD~245310) were acquired multiple times during each observing run.
The polarized standards are used to calibrate the polarization position angle.

%***Chris: below, do you mean ``custom-matured'' to ``custom-made''?
The raw data were bias subtracted and flatfielded in a standard way. Polarimetric reduction was carried out using custom-made IRAF scripts. All the $Q$ and $U$ spectra
obtained during a given observing run were coadded to improve the signal-to-noise ratio (S/N).

We calculate $P$ and $\theta$ from $Q$ and $U$ using the debiasing prescription of \citet{Wardle1974}:
\begin{align}
P =& \pm \sqrt[]{|Q^2 + U^2 - \tfrac{1}{2}(\sigma_Q^2 + \sigma_U^2)|},\\
\theta =& \tfrac{1}{2} \tan ^{ - 1} (U/Q).
\end{align}
The sign of P is chosen based on the sign of the result in the absolute-value operator.

\begin{figure*}
  \centering
  \includegraphics[width=0.95\textwidth, height=0.95\textheight, keepaspectratio]{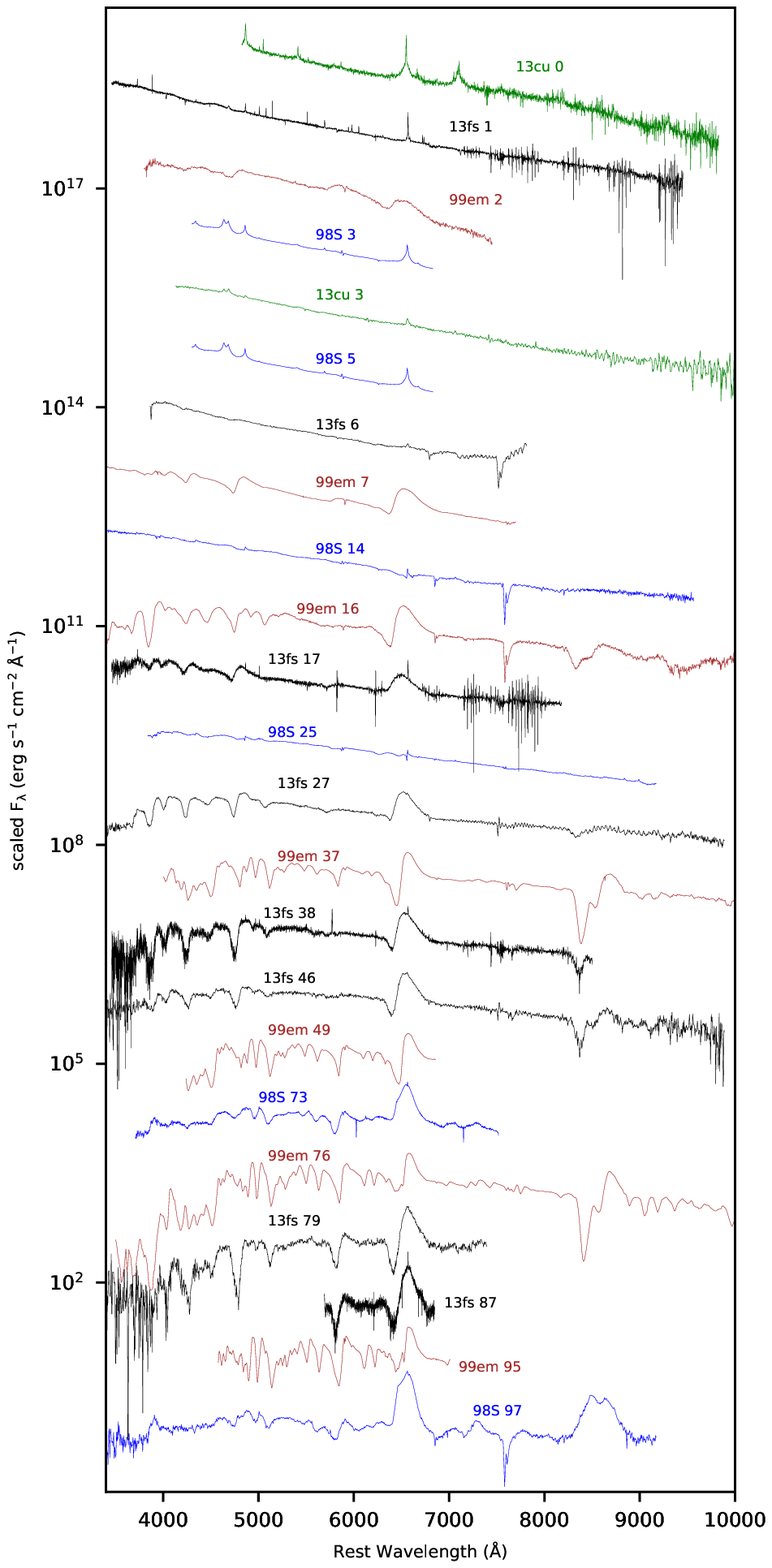}
  \caption{Spectra of SN~2013fs (red) compared with SN~1998S (blue), SN~1999em (brown), and SN~2013cu (green). Comparison spectra were retrieved from WISeREP. 
    The spectra for SN~2013cu originate from \citet{GalYam2014}; SN~1998S from \citet{Leonard2000}, \citet{Fassia2001}, and \citet{Fransson2005}; and SN~1999em from
    \citet{Leonard2002} and \citet{Hamuy2001}.}
  \label{fig:fscompspec}
\end{figure*}

\begin{figure}
  \centering
  \includegraphics[width=\columnwidth]{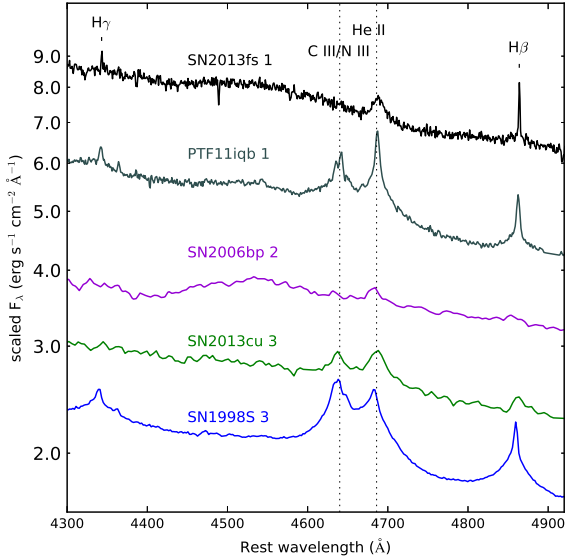}
  \caption{Comparison of the He\,\textsc{ii} $\lambda$4686 and C\,\textsc{iii}/N\,\textsc{iii} $\lambda$4640 lines in SN~2013fs (black) with those of SN~1998S (blue), SN~2006bp
    (purple), PTF11iqb (grey), and SN~2013cu (green). The rest-frame line centres are marked by the dashed vertical lines. The spectrum of SN~2006bp is
    from \citet{Quimby2007} and that of PTF11iqb from \citet{Smith2015}, retrieved from WISeREP. All spectra have been corrected for Milky
  Way Reddening.}
  \label{fig:HeIIcomp}
\end{figure}

\begin{figure}
  \centering
  \includegraphics[width=\columnwidth]{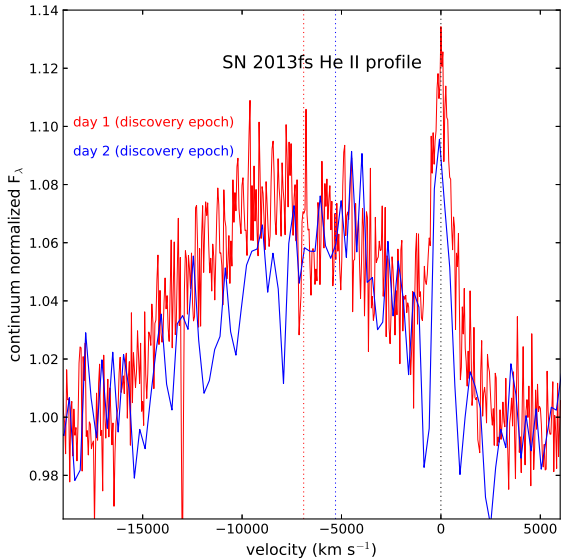}
  \caption{He\,\textsc{ii} line profile in SN~2013fs on days 1 and 2, with the zero point on the narrow component. The dashed vertical lines indicate the line
    centers of the components. They have been corrected for Milky Way Reddening with $E(B-V)$ = 0.0347~mag.}
  \label{fig:broadHeII}
\end{figure}

\begin{figure*}
  \centering
  \begin{minipage}{0.45\textwidth}
    \includegraphics[width=\columnwidth]{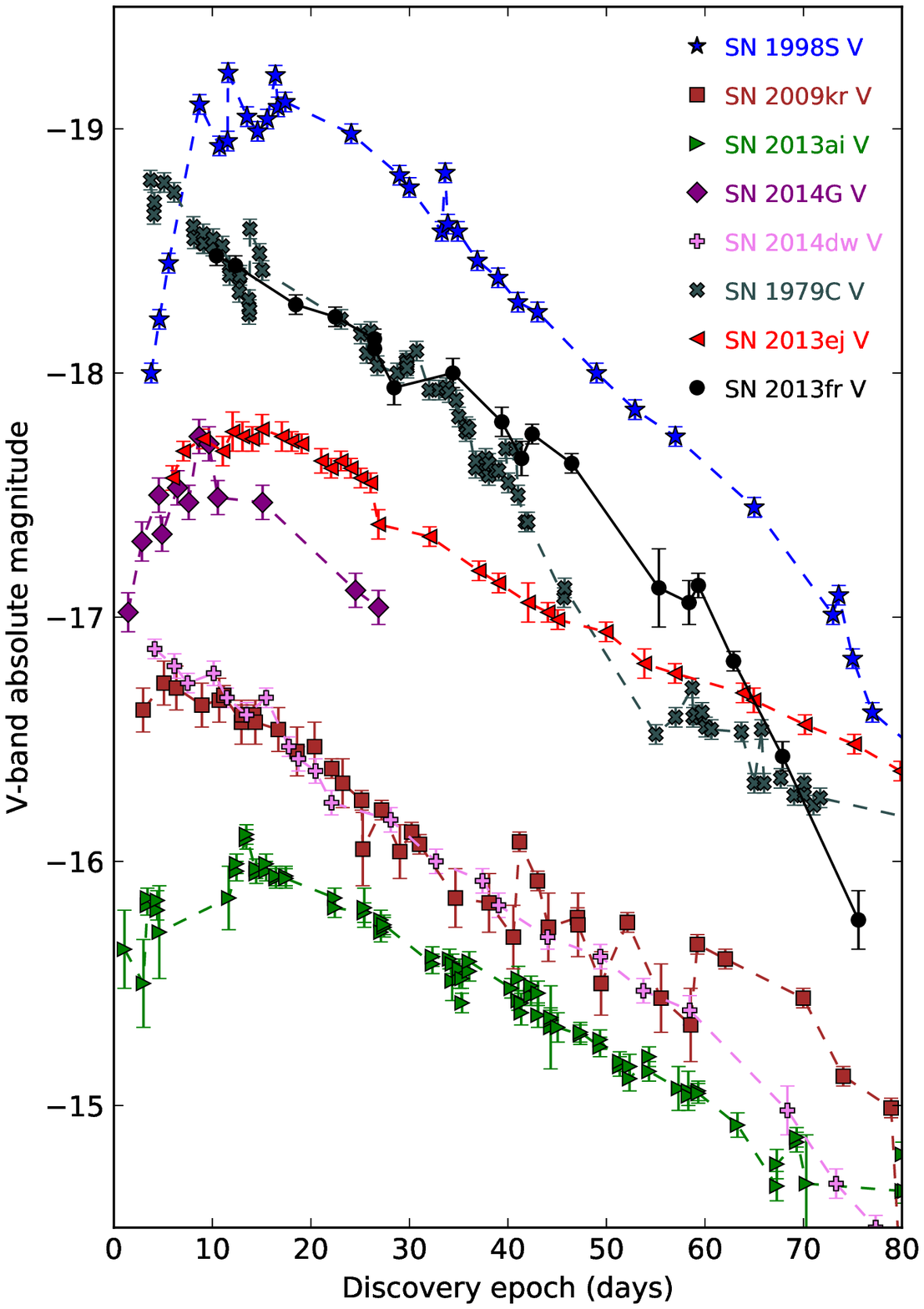}
    \caption{{\it V}-band comparison of SN~2013fr (black) with SN~1998S (blue), SN~2009kr (brown), SN~2013ai (green), SN~2014G (purple), SN~2014dw (pink), SN~1979C
      (blue-grey), and SN~2013ej (red). All photometry was obtained from the OSC, and has been corrected for Milky Way but not local reddening (if any). SN~2013ai,
      SN~2014G, and SN~2014dw all come from \citet{Valenti2016}, SN~1979C from \citet{Barbon1982}, SN~2009kr from \citet{EliasRosa2010}, and SN~2013ej from \citet{Huang2015}.}
  \label{fig:comparefrIIL}
  \end{minipage}%
  \hfill
  \begin{minipage}{0.45\textwidth}
    \includegraphics[width=\columnwidth]{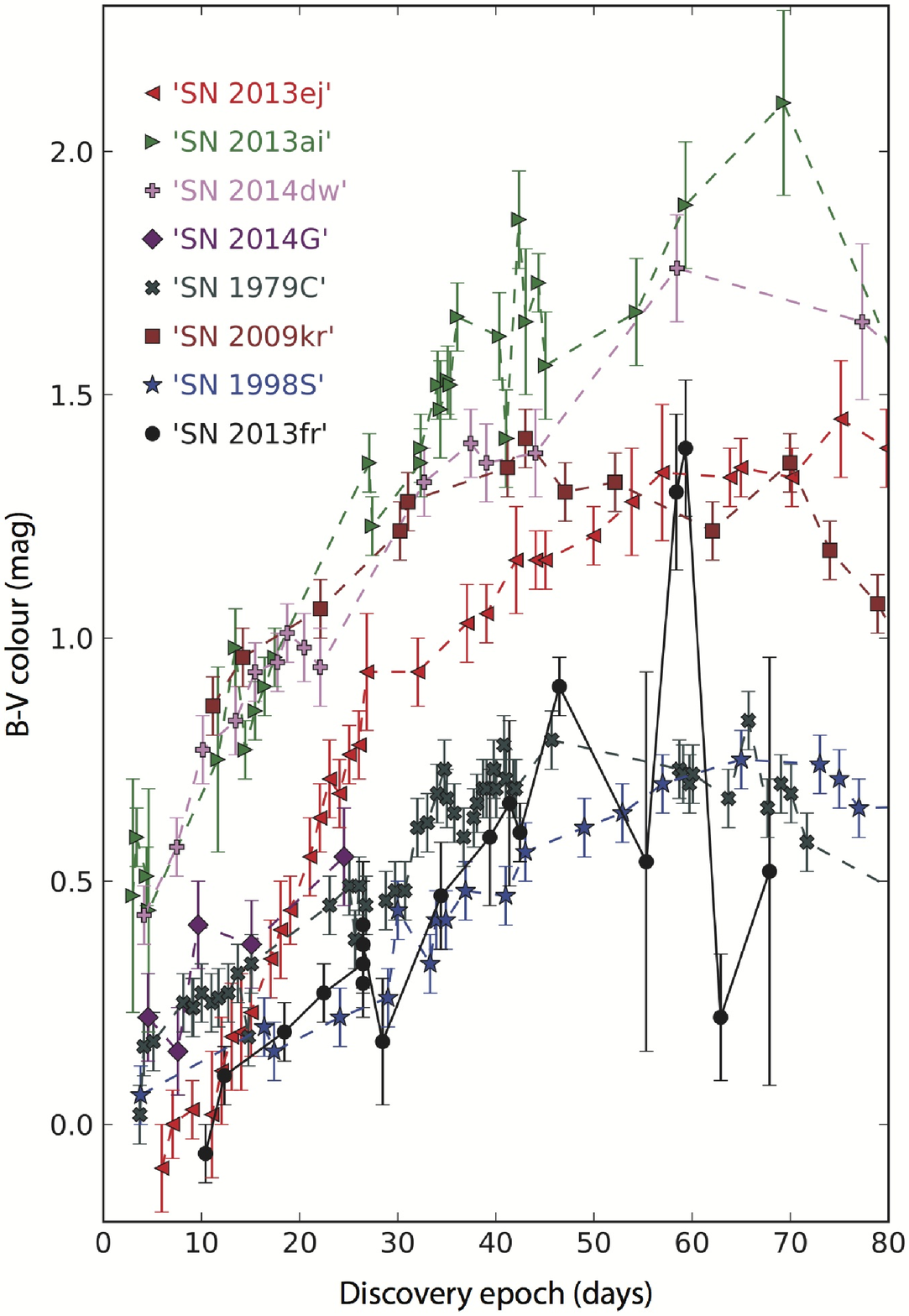}
    \caption{{\it B--V} colour evolution of SN~2013fr, SN~1998S, SN~2013ej, and the sample of SNe~II-L.}
    \label{fig:frcolorcompareIIL}
  \end{minipage}
\end{figure*}

\subsection{SN~2013fr} \label{obs:13fr}

\subsubsection{RATIR photometry} \label{obs:RATIR}
We obtained 18 epochs (days 7--76) of photometry of SN~2013fr using the multi-channel Reionization And Transients InfraRed camera \citep[RATIR;][]{butler12} mounted on
the 1.5~m Johnson telescope at the Mexican Observatorio Astrono\'mico Nacional on Sierra San Pedro M\'artir in Baja California, M\'exico \citep{watson12}. Typical
observations include a series of 80~s exposures in the {\it $ri$} bands and 60~s exposures in the {\it $ZYJH$} bands, with dithering between exposures. Given the lack of a
cold shutter in RATIR's design, infrared (IR) dark frames are not available.  Laboratory testing, however, confirms that the dark current is negligible in both IR detectors \citep{fox12}.

The data were reduced, coadded, and analysed using standard CCD and IR processing techniques in IDL and Python, utilising online astrometry programs {\tt SExtractor}
and {\tt SWarp}\footnote{SExtractor and SWarp can be accessed from http://www.astromatic.net/software.}. The photometry in the Sloan Digital Sky Survey (SDSS) {\it r} filter for the nights between
day 27 and 59 for which there is overlap with the KAIT/Nickel Johnson-Cousins {\it R} photometry show a slightly smaller rate of decline in the former. These data are given in
Table \ref{tab:RATIR}.

\subsubsection{DCT photometry} \label{obs:DCT}
One epoch of SN~2013fr photometry was obtained on day 139 using the Large Monolithic Imager (LMI) mounted on the 4.3~m Discovery Channel Telescope, in the SDSS  {\it g}, {\it r},
and {\it i} filters; no source was visible down to 23.29, 23.80, and 23.04~mag (respectively). The photometry was obtained in the same way as
the Kuiper photometry, and these deep upper limits set strong limits on the later light-curve decline rate (and possible plateau drop), as discussed in
Sections \ref{res:frphot} and \ref{2013fr}.

\subsubsection{Spectroscopy} \label{obs:frspec}
Seven low-resolution spectra were obtained using the Kast spectrograph on the 3~m Shane telescope at Lick Observatory \citep{Miller1993}, and one low-resolution spectrum was
obtained on day 46 using the IMACS spectrograph on the Baade Magellan telescope at Las Campanas Observatory \citep{Dressler2011} with the 300~line~mm$^{-1}$ grating. All
spectra are from the first 69 days post-explosion, and have been corrected for redshift using the {\sc DOPCOR} package in IRAF, and dereddened assuming $R_{V} = 3.1$. The
spectral evolution in the first 69 days is the subject of Section \ref{res:frevol}, and technical information on the instruments used to obtain these data is given in the bottom half of
Table \ref{tab:spectra}.

\subsection{Archival data} \label{obs:archive}

\subsubsection{Archival photometry} \label{obs:archphot}
The KAIT, {\it Swift}, and Kuiper photometry of SN~2013fs was supplemented with {\it r} and {\it R} band photometry obtained from the Open Supernova Catalogue (OSC) 
\citep{Guillochon2016}\footnote{https://sne.space}. The {\it r} and {\it U}-band data, as well as all of the {\it BVRI} photometry after day 49 post-discovery, were obtained from
the Las Cumbres Observatory Global Telescope Network (which observed SN~2013fs for 39 epochs covering days 0 to 106 after discovery. The {\it R}-band data come from the PTF48
telescope at Mount Palomar for the first week post-explosion. All archival photometry was added to the OSC by \citet{Valenti2016} after use in their paper exploring
correlations between SN~II diversity and possible progenitors.

\subsubsection{Archival spectroscopy} \label{obs:archspec}
The spectra obtained by Lick/Kast, Bok, Kuiper, Magellan, and MMT were supplemented by data obtained from online archives. For SN~2013fs, we obtained seven low-resolution spectra covering days 2 through 80, and four moderate-resolution spectra from days 1 to 52 via WISeREP \citep{WiSEREP}. The low-resolution spectra come from the
Public ESO Spectroscopic Survey of Transient Objects (PESSTO; \citet{PESSTO}), who obtained them using the Faint Object Spectrograph on the European Southern
Observatory (ESO) New Technology Telescope (NTT). The PESSTO spectra at days 27, 35, 46, and 53, originally split into two spectra, have been merged around 5500~\AA\
(the [O\,\textsc{i}] night-sky emission line was removed from these spectra). The moderate-resolution spectra come from the Australian National University (ANU) Wide-Field
Spectrograph (WiFeS) SuperNovA Program (AWSNAP) \citep{Child2016}. These spectra were originally used to reclassify SN~2013fs as a SN~II-P \citep{ATEL}. The first WiFeS
spectrum, taken within 48~hr of the initial explosion, is one of the earliest spectra obtained of a young SN, and is a direct probe of the CSM very close to the
explosion site; this is discussed in Section \ref{2013fs}.

For SN~2013fr, a single, day 4 PESSTO spectrum with a resolution of 5.4~\AA\ was obtained from WISeREP.

%% Analysis of light curves
\section{Results}\label{res}

\subsection{SN 2013fs}\label{res:13fs}

\subsubsection{Pre-discovery photometry \& {\it Swift} UVOT light curves}\label{res:uv}
SN~2013fs was discovered extremely young, within 48~hr of the explosion based on the time between the last KAIT upper limit and the discovery by \citet{CBET3671}, and was
detected within 8~hr based on the pre-discovery {\it R}-band photometry the day before (making SN 2013fs one of the only SNe detected so soon after explosion). As the initial rise to maximum is often proportional to $t^{2}$ \citet{Smith2015}, we fit the early-time {\it R}-band light curve (in magnitudes) to a quadratic function of the form $a(t-t_{0})^{2}$ (plotted as the dashed line in Figure \ref{fig:fsphot}). The fit suggests that the explosion occurred 1.5~days before
discovery, on 2013 Oct. 5.946 (MJD 56570.946) ($\pm 0.2$ days), consistent with the KAIT upper limits. Follow-up observations began almost immediately with both KAIT and
{\it Swift} UVOT. All photometry of SN~2013fs, including the eight most recent pre-discovery upper limits, is plotted in Figure \ref{fig:fsphot}; owing to the lack of a
resolved sodium doublet in the spectra, host-galaxy reddening is presumed to be negligible. The UV light curve of SN~2013fs is similar to the typical light curves of some SNe~IIn
and most SNe~II-P, rapidly fading linearly in the {\it UVW2/UVM2/UVW1} filters, and more slowly (but still linearly) in the {\it u} filter. The {\it b} and {\it v} filters
trace the KAIT {\it B} and {\it V} filters fairly well, though brighter by $< 0.1$~mag after the KAIT magnitudes were corrected for extinction.

Unlike many other SNe, the fast follow-up observations of SN~2013fs managed to catch a small rise in the {\it UVM2} and {\it UVW1} filters: $-0.185$ and $-0.249$~mag,
respectively. Over this period, the {\it UVW2} filter recorded only a 0.029~mag decline, suggesting that its maximum was reached very quickly, within 24~hr of the
explosion (by contrast, the {\it UVM2} and {\it UVW1} bands peak around 70~hr post-explosion). The maxima in the {\it UVW2/UVM2/UVW1} bands range from $-19.14$ to $-19.35$,
and the average decline in those bands is 0.23/0.24/0.17 mag~day$^{-1}$, consistent with the mean decline rates for SNe~II-P \citep{Pritchard2013}.

Comparison with the UV light curves of several SNe from \citet{Pritchard2013} (in this analysis, we look solely at the subset of SNe II-P/IIn
featured in Figure 6 of \citet{Pritchard2013}, as only this subset of their sample has derived explosion dates; we ignore SN~2008am and
SN~2008es, as these are SLSNe) and the SN~II-P SN~2014cx \citet{Huang2016} show that SN~2013fs is very similar to other SNe~II-P at similar
epochs, though brighter than average in the {\it UVW2/UVM2/UVW1} filters. The Milky Way reddening ($E(B-V)$) for the full sample is less
than 0.06 in all but one case (it is 0.09 for SN~2014cx). SN~2013fs is brighter in the {\it UVW2/UVM2/UVW1} filters than every other SN~II-P
in our sample at early times, usually by more than a magnitude. The only brighter SN is SN~2007pk, the sole SN~IIn in \citet{Pritchard2013}
the sample. No host reddening is available for most of these SNe, and so this excess brightness may be the result of higher host reddening
in the sample SNe. SN~2006bp in particular, has a very high host reddening of $E(B-V) = 0.37$~mag (from Table 2 in \citealt{Pritchard2013}).
This is noteworthy because SN~2006bp is the only SNe II-P in sample set known to have had WR-like emission lines at early times, and at a
difference of ~2.7 mag in the {\it UVW2} filter, it is the second dimmest SN II-P in the sample set.

The {\it UV--v} colour curves of SN~2013fs plotted in Figure \ref{fig:fscolors} behave like those of a typical SN~II-P; they start out
extremely blue, and rapidly redden with time. Most SNe~II-P settle into a UV plateau around 20 days after $v$-band maximum, and the reddening
levels off. This may be beginning to happen in the {\it UVW1--v} colour on the last epoch of UVOT observations, but it is not seen in
{\it UVW2} and {\it UVM2}. It would be tempting to say that this indicates a deviation from the typical UV colour evolution of a SN~II-P,
were it not for the fact that at late times host contamination in the deeper UV filters is not negligible and can cause up to a 1~mag
difference in the photometry.

\subsubsection{Optical light curves}\label{res:fsoptical}
As with the UV photometry, SN~2013fs peaks in the optical very quickly after discovery, reaching a {\it B}-band maximum $M_{B} = -17.66$~mag and a {\it V}-band maximum
$M_{V} = -17.69$~mag on days 4 and 6 after the discovery. After peak, it settles into the plateau phase with a luminosity around $M_{R} = -17.50$~mag, where it
remains for at least 50 days. The unfiltered, {\it r}, and {\it BVRI} photometry show that the plateau drops off around day 75, dimming by more than a magnitude in the {\it r/R}
filters over the next 34 days, after which our photometric coverage ends.

SN~2013fs exhibits a shorter plateau relative to many SNe~II-P (by about a month), and a brief, sharp peak in the early light curve not seen in most SNe~II-P. Detailed comparison of SN~2013fs with the interacting Type IIn SN~1998S, the gap-bridging SN~IIn PTF11iqb, and the archetypal Type II-P SN~1999em is the subject of Section \ref{2013fs}.

\subsubsection{Spectral evolution}\label{res:fsspecevol}
SN~2013fs was initially classified as a young SN~IIn, owing to the narrow H$\alpha$ emission in the first epoch. The day 1 spectrum in Figure \ref{fig:day1} exhibits a blue
continuum with narrow lines, with hydrogen emission features and He\,\textsc{ii} $\lambda$4686 having a broad blueshifted base with FWHM $\approx$ 11,500 km\ s$^{-1}$, and a
narrower component with Lorentzian wings and FWHM $\approx$ 1000 km\ s$^{-1}$. This has been observed in a handful of SNe for which very early spectra are available,
including SN~1998S \citep{Shivvers2015}, SN~2006bp \citep{Quimby2007}, SN~2013cu \citep{GalYam2014}, and PTF11iqb \citep{Smith2015}. Similar to SN~2006bp and dissimilar to
SN~1998S, SN~2013cu, and PTF11iqb, only He\,\textsc{ii} is distinguishable, while the common C\,\textsc{iii}/N\,\textsc{iii} $\lambda$4640 line is either very weak or
absent. Deblending the emission line using Lorentzian fits shows that if there is a distinct emission line at 4640~{\AA}, then the C\,\textsc{iii}/N\,\textsc{iii} line is
weaker than He\,\textsc{ii}; the equivalent width (EW) ratio EW($\lambda$4640)/EW($\lambda$4686) $\lesssim 0.08$.

All spectra of SN~2013fs are plotted in Figure \ref{fig:fsallspec}. The narrow H$\alpha$ persists for at least a week post-discovery. The line profile broadens by day
17 and develops a standard P Cygni profile, typical of normal SNe~II-P. The moderate-resolution spectra indicate that a small narrow component remains, and other narrow
features in Figure \ref{fig:day1} are indicative of an H\,\textsc{ii} region. Using the moderate-resolution MMT spectrum and WiFeS spectra, the H$\alpha$ line profile of
SN~2013fs is plotted on the left side of Figure \ref{fig:halpha}. The H$\alpha$ profile on day 1 has faint [N\,\textsc{ii}] emission. These features fade as the P Cygni
profile develops, and the velocity of the absorption trough becomes increasingly redshifted from day 17 to 87. After the plateau drop, the P Cygni absorption becomes
narrower, with FWHM $\approx$ 3900 km\ s$^{-1}$.

\subsubsection{Spectropolarimetric evolution}\label{res:specpol}

Establishing an accurate interstellar polarization (ISP) is often necessary to properly interpret the polarimetric results. Here, we determine an ISP value by assuming the
intrinsic polarization of our Epoch-2 data (i.e., the data with the highest S/N; days 21 and 23 post-discovery) is zero or centred around $(q,u) = (0,0)$. This
assumption may be incorrect, but a different ISP value will not alter our interpretation significantly. Therefore, we adopt the inverse-variance weighted averages over the
full spectra for Epoch~2 of $(q,u) = (-0.45,-0.26)$ as the ISP value. All the figures show only ISP-subtracted data. 

Figures \ref{fig:specpol1} through \ref{fig:specpol4} show the scaled $F_{\lambda}$, and normalised linear Stokes parameters $q$ and $u$ as a function of wavelength for the
four epochs as specified in Section \ref{obs:specpol}. The unbinned $q$ and $u$ data are shown in light red and data binned by 15 pixels is overplotted in blue. The flux data
are always unbinned. Figure \ref{fig:quplane} shows the binned data from all four epochs plotted in the ($q,u$) plane colour coded by wavelength. The low-S/N data
below 4200~{\AA} are not plotted. The light-grey crosshairs show the position of the origin before ISP subtraction. Some previous studies have chosen an ISP value such that
the data along a preferred axes do not pass through the origin in the ($q,u$) plane.  This eliminates a $90^\circ$ rotation at some point in the spectrum. However, wavelength-dependent optical-depth effects and the specific source-scatterer geometry can result in a $90^\circ$ rotation at any wavelength \citep{Dessart2011}; hence, we are not constrained by that requirement.

There is some deviation from the average in both $(q,u)$ and $(P,\theta)$ across the H$\alpha$ absorption line in all epochs, but it is most pronounced in Epoch 3 (days 52 and 57), when the absorption feature is largest. This is perhaps easiest to see in the ($q,u$) plots in Figure \ref{fig:quplane}, where the data near the H$\alpha$
absorption line are coloured yellow-green. This wavelength-dependent variation demonstrates that there is intrinsic SN polarization of $\sim 1$\%.

There appears to be a preferred axis for the continuum data in Epoch 1 that runs through the origin at approximately $50^\circ$ and $230^\circ$ in the ($q,u$) plane or $25^\circ$
and $115^\circ$ on the sky. In Epoch 3, that preferred axis appears to have rotated by about $90^\circ$, to $140^\circ$ and $320^\circ$ in the ($q,u$) plane or $70^\circ$ and
$160^\circ$ on the sky. The data in Epoch 2 are tightly clustered around the origin, perhaps in transition from one preferred axis to the other.  In Epochs 1 and 3,
the H$\alpha$ line deviates from the continuum orthogonal to the preferred axis. H$\alpha$ in Epoch 1 is along the preferred axis for the continuum of Epoch 3 and
H$\alpha$ in Epoch 3 is along the preferred axis for the continuum of Epoch 1. 

This polarimetric evolution, with a rotation of the preferred axis, isn't necessarily a result of a change in the geometry of just the scattering envelope but may be a result
of a change in the source-scatterer geometry as the photosphere recedes through the ejecta. Total continuum polarization is small (< 1{\%} at all epochs, and < 0.5{\%} at
Epoch 2, where the S/N is highest. This mild asymmetry is similar to that of SN~1998S \citep{Leonard2000,Shivvers2015}.

\begin{figure*}
  \centering
  \includegraphics[width=0.95\textwidth, height=0.95\textheight, keepaspectratio]{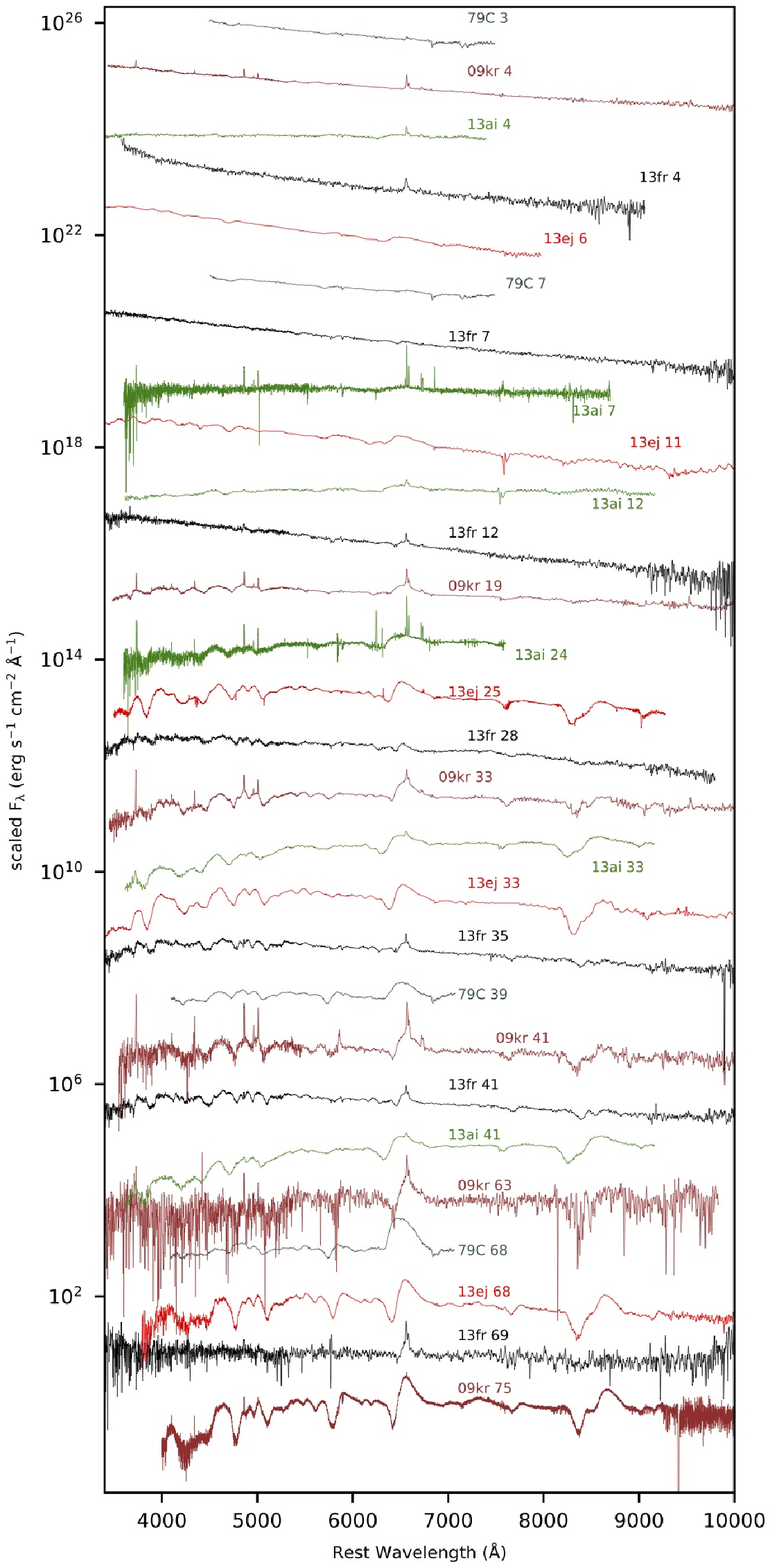}
  \caption{Spectra of SN~2013fr (black), compared with SN~2013ej (red), SN~1979C (grey), SN~2013ai (green), and SN~2009kr (brown). Spectral comparisons with SN~2014G and
    SN~2014dw were not possible owing to the lack of public spectra. All spectra with the exception of that of SN~2009kr were retrieved from WISeREP; the spectra of SN~2009kr were obtained
    from Nancy Elias-Rosa via email. All dates are taken with respect to discovery, and all spectra have been corrected for Milky Way reddening.}
  \label{fig:frcompspecIIL}
\end{figure*}

\subsection{SN 2013fr}\label{res:13fr}

\subsubsection{Optical and IR photometry of SN~2013fr}\label{res:frphot}
All photometry of SN~2013fr is plotted in Figure \ref{fig:frphot}, and spectral energy distributions (SEDs) are shown in Figure \ref{fig:frSED}. In contrast to SN~2013fs, SN~2013fr was caught potentially
a week or more after explosion, and all optical bands are already declining by the first photometric epoch. This puts a lower limit on the true maxima of $M_{B} = -18.65$
mag and $M_{V} = -18.48$ mag. The decline is approximately linear in the {\it BVRI}-bands, though there is a break in the photometry at about
60 days, at which point the decay slope increases noticeably in the {\it VRI}-bands, perhaps in transition to the nebular phase, though the
limited temporal coverage makes this uncertain.

The slope is more shallow for the {\it YJH} filters than for the optical, with average decline slopes of 0.05/0.04/0.03 mag~day$^{-1}$ for the
{\it B/V/R} filters and 0.02/0.02/0.01 mag~day$^{-1}$ for the {\it Y/J/H} filters, respectively. The brightness in the {\it J} and {\it H} filters is almost constant until day
63 (at which point they begin a decline similar to the break seen at optical wavelengths). At the end of coverage on day 75, all filters except {\it H} have seen a decrease of more than a magnitude. The SEDs of SN~2013fr show that an IR excess appears around
day 42 and becomes prominent by day 75. The temperature decreases with time as expected, but at day 42 and thereafter the SED is well described by a double blackbody model.

After coverage ends on day 75, no point source is found in the {\it gri}-bands from the day 139 DCT photometry. Given a 3$\sigma$ limiting magnitude of $m_{r} = 23.8$~mag,
SN~2013fr has an average decline rate exceeding 0.08 mag~day$^{-1}$ in the {\it r}-band, noticeably higher than the 0.03 mag~day$^{-1}$ average {\it r}-band decline rate in
the first 75 days, pointing to an accelerated decline after the end of coverage. This decline rate is consistent with, if somewhat elevated above, the later-time
evolution of SNe~II-L \citep{Faran2014}. Unfortunately, no photometry was obtained during the nebular phase of SN~2013fr that would allow us to obtain a good estimate of
the $^{56}$Ni mass. SN~2013fr is compared with several SNe~II-L in Section \ref{2013fr}.

\subsubsection{Spectral evolution of SN~2013fr}\label{res:frevol}
The spectra of SN~2013fr are plotted in Figure \ref{fig:frallspec}, and the H$\alpha$ line profile is shown in the right half of Figure \ref{fig:halpha}. The spectrum
reddens continually with age, with relatively strong line blanketing appearing on day 28 and onward. In the final spectrum on day 69, the spectrum is dominated by the
H$\alpha$ feature, but this spectrum is noisy and most of the major SN features are unresolved. The continuum slope of the spectra changes slowly, and remains relatively blue
until late times, although the apparent colour gets redder because of increased line absorption.

The day 4 spectrum of SN~2013fr is similar to that of SN~2013fs, and may contain narrow H$\alpha$ emission attributable to the SN. The H$\alpha$ feature develops an
intermediate-width P Cygni profile and the narrow lines disappear within a few days.

The remaining narrow lines in the later-time spectra resemble those of the host, and this feature is likely to be contamination from a neighbouring H\,\textsc{ii} region. In the
first spectrum on day 4, H$\alpha$ has EW = $-24.11$~{\AA}, while the narrow H$\alpha$ feature at later times has EW $> -7$~{\AA} for all but the final
epoch. Using the {\it r}-band photometry from day 12 as representative of the continuum $F_{\lambda}$ on day 12, and interpolating back to day 4, the ratio of the narrow
H$\alpha$ line flux (EW $\times F_{\lambda}$) on day 4 to that on day 12 is $\sim 6.5$. Host contamination may not be able to account for the much stronger narrow
H$\alpha$ emission on day 4. The P Cygni absorption velocity is variable with time, in the range 4800-5200 km\ s$^{-1}$.

\section{SN~2013\lowercase{fs} and its place among SN\lowercase{e} II-P \& II\lowercase{n}}\label{2013fs}
During the rise to maximum light, SN~2013fs is a close match photometrically to PTF11iqb, which is a very close spectroscopic analogue of
SN~1998S \citep{Smith2015}. As seen in Figure \ref{fig:comparefs}, SN~2013fs has an absolute magnitude between those of SN~1999em and
SN~1998S. It most closely matches PTF11iqb, but with a faster rise to maximum and a shorter, fainter plateau. The rapid rise to peak
brightness is suggestive of a progenitor with an extended envelope, like a RSG. Such a progenitor was also proposed for SN~1999em, SN~1998S,
and PTF11iqb; the main difference between these explosions lies in their inferred pre-SN mass loss (see \citealt{Morozova2016} and
\citealt{Smith2015}).

The colour evolution of SN~2013fs (plotted in Figure \ref{fig:fscolorcompare}) tells a similar story, falling in the range between those of
normal SNe~II-P like SN~1999em and SNe~IIn like SN~1998S. Its {\it B--V} evolution reddens quickly in the early days when the SN~IIn features
are still present, and after day 10 begins to slow. The {\it B--V} colour of SN~2013fs never becomes quite as red as SN~1999em, but it
remains redder than either SN~1998S or PTF11iqb \citep{Smith2015}. This indicates that the blue excess, which mostly disappears by day 20, is
correlated with stronger CSM interaction. After the plateau drop, the {\it B--V} colour evolution levels off. This occurs at about the same
time in SN~1998S and SN~1999em, with the {\it B--V} colour of SN~2013fs remaining much closer to that of SN~1999em, consistent with the lack
of CSM interaction signatures at late times.

In the redder filters, SN~2013fs mirrors SN~1999em throughout the plateau phase. After the plateau drop, the {\it V--R} and {\it V--I}
colours of SN~2013fs become very red ($\sim 0.3$~mag over that of SN~1999em in {\it V--R}, and $\sim 0.4$~mag over in {\it V--I}), matching
SN~1998S at similar epochs.

Spectral comparisons between SN~2013fs and other SNe at early times are shown at the top of Figure \ref{fig:fscompspec}. SN~2013fs is
initially spectroscopically distinct from SN~1999em, but begins to increasingly resemble it as the plateau phase continues, and the P Cygni
absorption deepens. Based on pre-explosion imaging, \citet{Smartt2002} suggested that the progenitor of SN~1999em was likely to be a RSG
with an upper mass limit of $M \approx 12$~M$_{\odot}$. The plateau phase was shorter for SN~2013fs than it was for SN~1999em by more than a
month, which usually indicates a lower-mass H envelope, and SN~2013fs has a greater decline in the {\it R} band during the plateau (SN~1999em,
by contrast, has nearly constant {\it R}-band magnitude during the plateau). SN~2013fs is also almost a magnitude brighter in {\it R} for the
entire duration of the plateau, conforming to the inverse correlation between plateau decline and high peak brightness found by
\citet{Anderson2014}.

\citet{Smith2015} argued that the spectroscopic signatures of SNe~IIn could extend to less extreme mass-loss scales or shorter durations, and
that such SNe would be recognized as Type IIn only if discovered very young. SN~2013fs was discovered within 8~hr of explosion, and is an
illustrative example of this phenomenon. The narrow Balmer lines produced by CSM interaction are prominent during the rise to maximum light,
but quickly fade over the first two weeks, giving way to broad P Cygni profiles. The last three spectra in Figure \ref{fig:fsallspec} show
that the broad H$\alpha$ component is slowly narrowing again at late times. The short duration of the narrow H$\alpha$ emission may suggest
strong mass loss for a limited time before explosion. This conclusion is supported by the concurrent bump in the early light curve (modeled
for SN~2013fs by \citealt{Morozova2016}), and by other emission lines suggestive of CSM interaction. Since the narrow emission disappears
quickly, the CSM created by this mass loss is relatively compact.

Most of the narrow features in the early-time spectrum are attributable to H\,\textsc{ii} region emission, but the multi-component
He\,\textsc{ii} $\lambda$4686 feature is more interesting. The lack of other WR-like lines and the rapid fading indicates that these lines
are direct probes of the CSM extremely close to the site of the explosion. The He\,\textsc{ii} line has EW($\lambda$4686) = 3.018~{\AA} on
day 1 ($\sim 2.5$ days post-explosion) and 1.588~{\AA} on day 2.58 ($\sim 4$ days post-explosion), and is unresolved entirely in the day 4
spectrum. The He\,\textsc{ii} flux ratio $f(2.5)/f(4) \approx 2$, following the EW ratio. Like other SNe with WR-like early features
\citep{Khazov2016}, SN~2013fs has a maximum luminosity $M_{R} > -17.6$ mag.

A comparison of the He\,\textsc{ii} $\lambda$4686 and C\,\textsc{iii}/N\,\textsc{iii} $\lambda$4640 lines in SN~2013fs, SN~1998S, SN~2013cu,
SN~2006bp, and PTF11iqb is shown in Figure \ref{fig:HeIIcomp}. These lines typically fade within 72~hr of the first spectral epoch, as was
the case in SN~2013fs. They might arise due to ionisation from the UV flash resulting from shock breakout \citep{Khazov2016}, or as the
result of CSM interaction (e.g., X-rays from very early shock interaction; see \citet{Smith2015}). The He\,\textsc{ii} emission also shows a
broad, blueshifted component (plotted in Figure \ref{fig:broadHeII}). This was also seen in SN~2006bp and to a lesser extent PTF11iqb, both
of which also had plateau light curves. This component arises from the SN ejecta beneath the CSM shell, and both the broad and narrow
He\,\textsc{ii} vanish by day 4 (day 5.37 post-explosion), owing to shock cooling as the ejecta expand. The CSM remains important for at
least a few days after these high-ionisation lines have disappeared, as indicated by the pre-plateau bump in the light curve.

Following the method of \citet{Quimby2007}, it is possible to obtain an upper limit on the maximum distance of the shell from the SN
photosphere at the moment of SBO. Based on fits to the P Cygni absorption minimum of H$\beta$ on day 17 and of H$\alpha$ on day 21, the
expansion velocity of the ejecta is likely between 8400 (H$\alpha$) and 9300 (H$\beta$) km~s$^{-1}$. These later measurements are consistent
with a typical early-time expansion velocity of $\sim 10,000$ km~s$^{-1}$, and the blue end of the broad He\,\textsc{ii} component shows that
some components of the shock underneath the CSM have expansion velocities $\gtrsim 15,000$ km~s$^{-1}$. The H$\alpha$ line ceases to be narrow
some time after day 6, and the light curve enters the plateau phase around three weeks later, after which SN 2013fs is photometrically and
spectroscopically a normal SN II-P.

Assuming an average expansion velocity $\bar{v}_{\rm ejecta} \approx 10,000$ km~s$^{-1}$ during the initial 7.5 days post-explosion, we find the
CSM to be confined within $6.5 \times 10^{14}$~cm ($\sim 43$~au from the progenitor photosphere at SBO). If the circumstellar He\,\textsc{ii}
is ionised in the early-phase shock interaction, then assuming it began close to the time of our first spectrum (2.5 days post-explosion) and
that the approaching ejecta below the CSM at these early phases expanded at $\sim 15,000$ km~s$^{-1}$ (as measured from the blue edge of the
broad He\,\textsc{ii}), then the CSM shell must mostly lie beyond $3.24 \times 10^{14}$~cm.

The broad He\,\textsc{ii} feature is absent in SN~1998S, SN~2013cu, and PTF11iqb. Since the broad feature decays quickly, its absence in
these SNe may be the result of differing discovery times, or it may be blended with the strong C\,\textsc{iii}/N\,\textsc{iii} emission
present in their early-time spectra. Of all the known SNe which exhibit WR-like lines, only SN~2006bp is also seen to have both comparably
weak emission of the C\,\textsc{iii}/N\,\textsc{iii} $\lambda$4640 blend and a blueshifted He\,\textsc{ii} component. Interestingly,
SN~2006bp has a similarly short-duration plateau like SN~2013fs, with very similar limits on the extent of the CSM shell \citep{Quimby2007}.

The nature of the mass-loss that lead to this shell and the total mass shed have been the source of a great deal of recent work. There are
three recent papers which come to mostly different conclusions on one or both of these questions. \citet{Morozova2016} uses the SuperNova
Explosion Code (SNEC; \citealt{Morozova2015}) with the light curves from \citet{Valenti2016}, and assuming a smooth, wind-like density
profile (though they suggest the phenomenon responsible could be eruptive), infers a very large amount of mass lost in a very short
period of time (and consequently, a very high M$_{\odot}$).

\citet{Takashi2017} use a progenitor model and radiation hydrodynamics simulations
to explore the impact of wind acceleration on the duration and severity of the late stage mass-loss, finding a similar total ejected mass
to \citet{Morozova2016}, but that the phase may have lasted 5--500 times longer than \citet{Morozova2016} suggests, and that M$_{\odot}$ may be
as much as two orders of magnitude lower.

\citet{Yaron2017}, which infers an even tighter constraint on the explosion time, uses radiative transfer modeling and
extremely early time spectra to infer the pre-SN mass-loss rate. They argue for the importance of a "superwind" phase of late stage
mass-loss, and also raise the possibility of late-time eruptive events.

More recently, \citet{Dessart2017} find that mass-loading in the intermediate region between the progenitor and its normal radiation-driven
wind can fully account for the early signatures of interaction seen in SN~2013fs (and one of our comparison SNe, SN~2013cu), and that no
transient super-wind or eruptive mass-loss phases are required at all. These will all be discussed in turn.

\citet{Morozova2016} fit the photometry of SN~2013fs from \citet{Valenti2016} using a light curve model that invokes a dense CSM created by a
steady-state wind, assuming spherically symmetric conditions. They determine the most likely progenitor is a RSG with
$M_{\rm ZAMS} = 13.5$—14.0~M$_{\odot}$, and the explosion energy to be E$_{\rm fin} = 0.8$—1.0~$\times 10^{51}$~erg. Using wind speeds
v$_{\rm wind} =$ 10 (100) km~s$^{-1}$, they find the immediate pre-SN mass-loss rate of SN~2013fs to be
$\dot{M} = 0.15\ M_{\odot}$~yr$^{-1}$ ($\dot{M} = 1.5 M_{\odot}$~yr$^{-1}$), with an assumed outer radius of 2100-2300 R$_{\odot}$ (9.8-10.7 AU).
This is higher than any known RSG wind \citep{Smith2014}, and implies the duration of this strong wind before core-collapse is $\sim$ 5 (0.5)
years. In both cases, the total mass of the ejected CSM is around 0.4 M$_{\odot}$.

Such a high mass loss rate suggests a very late time eruptive event in the last few years (for a slower velocity outburst) or even months
(for a higher velocity outburst) before core collapse. Based on the wind velocities from Table 1 of \citet{Morozova2016} and the extent of
the CSM shell inferred from our photometry/spectra, the mass-loss event began within 20~yr of core-collapse if v$_{\rm wind} = 10$~km~s$^{-1}$,
or within 2~yr if v$_{\rm wind} = 100$~km~s$^{-1}$.

The mass-loss rate inferred by \citet{Yaron2017} uses the same wind velocities, but comes to much more modest values of $\dot{M}$. Using the
He\,\textsc{ii} $\lambda$4686 and H$\alpha$ line fluxes in conjunction with temperature fits obtained using the CMFGEN radiative transfer
code \citep{Hillier1998}, they find that $\dot{M}$ is best fit in the range $2-4 \times 10^{-3}$~M$_{\odot}$~yr$^{-1}$ (using the same wind
speeds as in \citet{Morozova2016}); dramatically lower than that inferred by \citet{Morozova2016}.

Their estimate of $\dot{M}$ scales with the wind velocity by $\frac{v_{\rm wind}}{100 km~s^{-1}}$, and so even with unrealistic estimates for the
wind speed (1000 km~s$^{-1}$ or higher), the value of $\dot{M}$ does not exceed about 10$^{-2}$ M$_{\odot}$~yr$^{-1}$. In addition to the early
time spectra, they are also able to user later radio non-detections to rule out some longer term phases of weaker mass-loss (in the range
$6 \times 10^{-6}-1 \times 10^{-3}$~M$_{\odot}$~yr$^{-1}$). They do not calculate the total mass of the ejected CSM explicitly, though their
calculated mass-loss rate and very short preferred timescale ($\sim$ 100 days prior to core collapse) argues for a CSM mass not much more than
0.01 M$_{\odot}$.

Both \citet{Yaron2017} and \citet{Morozova2016} assume constant wind velocities, and the values of $\dot{M}$ and M$_{CSM}$ in each differ
considerably. \citet{Takashi2017} drop the assumption of constant velocity, instead modeling the wind with a $\beta$-law profile of the form
$v_{\rm wind}(r) = v_{0} + (v_{\infty}-v_{0}(0))(1-\frac{R_{0}}{r})^{\beta}$, with R$_{0}$ being the surface of the progenitor. Such wind acceleration
models have been used for galactic RSGs in the past, and typically find $\beta > 3$. The same is true here, and they find $\beta \approx 5$
provides the best fit to the early time {\it r}-band light curve.

Their model assumes that v$_{\infty}$ = 10 km~s$^{-1}$, and they do not provide fits for higher values of v$_{\rm wind}$. However, even for this
slower wind velocity, they find that the wind acceleration model predicts that the compact CSM around SN~2013fs was lost by the progenitor
within 500 years of core collapse, and possibly as little as 100 years. The total CSM mass is very similar to that found in
\citet{Morozova2016}. For shorter mass-loss durations, like those in \citet{Morozova2016} and \citet{Yaron2017}, the wind acceleration model
predicts that r$_{CSM} < 10^{14}$~cm, too compact to recreate the WR-like spectral features at early times. As a result, they favour a
super-wind phase starting no sooner than 550 years and no later than 100 years prior to core collapse \citep{Takashi2017}.

The most recent paper that discusses SN~2013fs, \citet{Dessart2017}, combines the observations of \citet{Yaron2017} with both radiative
transfer modeling and radiation hydrodynamics simulations (\citealt{Yaron2017}, in contrast, did the former but not the latter) to find a
good model of the wind/atmospheric density scale height and $\dot{M}$. The mass-loss values they used in their simulations range from
10$^{-6}$--10$^{-2}$ M$_{\odot}$~yr$^{-1}$, and scale heights range from 0.1--0.3 R$_{\star}$. Their value of v$_{\rm wind}$ at the outer edge of
the compact CSM is 50 km~s$^{-1}$, midway between the 10-100 km~s$^{-1}$ estimates used in the other papers.

Their primary interest is in the effect of the progenitor's outer atmosphere and circumstellar material just above R$_{\star}$ on the shock
breakout and early time spectra rather than mass-loss, but they propose an alternative not considered by any of the other papers. All
spatially resolved galactic RSGs have cocoons of material above the edge of the photosphere, and they propose that the compact CSM seen in
SN~2013fs is attributable to the same phenomenon. In that case, there is no need to invoke a late time super-wind or eruptive mass-loss
phase. Further, they challenge the "flash ionisation" interpretation of \citet{Yaron2017}, \citet{GalYam2014}, and \citet{Khazov2016},
arguing that the persistence of the WR-like lines is the result of the very high UV flux in the aftermath of the shock breakout, not the
initial UV flash.

We see the spectroscopic signatures of strong CSM interaction only for the first week or so after core-collapse, whereas the photometry, well
fit by the light curve models of \citet{Morozova2016} and \citet{Takashi2017}, shows that excess luminosity from continuing CSM interaction
lasts until around 20 days post-explosion. By the time the narrow lines disappear, a week or so post-explosion, the fastest ejecta
(v$_{\rm ejecta} \sim 10000$-15000~km~s$^{-1}$) will have expanded beyond the radius of the CSM shell, causing any remnant narrow emission from
continuing interaction to become obscured in the expanding ejecta. This scenario suggets asymmetry in the CSM, as was the case in PTF11iqb.
Unlike PTF11iqb, where the CSM interaction re-emerged during the nebular phase as a result of more distant CSM ($\sim 200$~AU;
\citealp{Smith2015}), the lack of post-plateau narrow emission in SN~2013fs suggests that CSM interaction is quenched after the excess
luminosity fades and the plateau phase begins.

A 0.5-100 year timescale would imply that even normal SNe~II-P from RSGs experience advanced mass loss in the final stages of nuclear burning,
producing SN~IIn spectra \citep{Smith2009,Smith2015}. The mass-loss phenomenon responsible could be a very strong, possibly accelerated
super-wind, or (if the higher value of $\dot{M} \approx 1.5$~M$_{\odot}$~yr$^{-1}$ found by \citealt{Morozova2016} or the 100 day timescale
preferred by \citealt{Yaron2017} are accurate) an eruptive event in the final stages of stellar evolution prior to core collapse
\citep{SmithArnett2014,Quataert2012}. If the mass-loss phenomenon is eruptive, then CSM velocities above 100 km~s$^{-1}$ are feasible, and the
event could have happened as little as a few years before core collapse).

Only a handful of SNe are known to exhibit the spectroscopic signatures of CSM interaction, but this is likely a discovery-time problem:
\citet{Khazov2016} find that they occur in as much as 18\% of young SNe~II, indicating that very late-stage mass-loss events may be common
occurrences. On the other hand, if \citet{Dessart2017} is correct and the compact CSM is nothing but an extended cocoon of surrounding
material ejected earlier on, like those seen in spatially resolved galactic RSGs, the phenomenon could be near universal, and the lack of
observed CSM interaction in most young SNe~II-P would be attributable to variations in the density of the compact CSM. In this case,
SN~2013fs and SN~2006bp represent RSGs with high density close CSM, with narrow interaction lines lasting as long as a few days
(\citealt{Dessart2017} suggests that these lines would last at best a few hours for RSGs with tenuous CSM).

\section{SN~2013\lowercase{fr}: a SN II-L with possible early-time CSM interaction and late-time dust}\label{2013fr}
There is no unambiguous plateau drop observed for SN~2013fr, and all filters decline at a relatively constant rate, with the {\it V} band in particular dropping by $\sim 1.3$ mag in the 45 days after our first $V$-band detection. The narrow lines disappear and reappear, but beyond
possibly the first epoch, these are most likely due to changes in seeing and host
contamination. Thus, we conclude that SN~2013fr is best described as a SN~II-L which may exhibit early CSM interaction. Adopting a $V$-band decline of more than 0.5~mag in the first 50 days as the defining feature of SNe~II-L from \citet{Faran2014}, it is a clear
SN~II-L. {\it V}-band photometric comparisons of SN~2013fr with SN~1998S, several SNe~II-L (SN~1979C, SN~2009kr, SN~2013ai, SN~2014G, SN~2014dw), and SN~2013ej (an intermediate
object between SNe~II-P and SNe~II-L; \citealp{Mauerhan2016}) is shown in Figure \ref{fig:comparefrIIL}. SN~2013fr has a light curve similar to that of SN~1979C at early
times (SN~1979C was discovered relatively young; this similarity suggests the same may be true of SN~2013fr), and it is brighter than all SNe in the sample other than SN~1979C and SN~1998S. The decline rate of SN~2013fr is consistent with all of the well-sampled SNe~II-L, and
is modestly steeper than that of SN~2013ej.

The {\it B--V} colour evolution of SN~2013fr, SN~2013ej, and the SN~II-L sample is plotted in Figure \ref{fig:frcolorcompareIIL}. The overall
colour evolution of SN~2013fr is similar to the other SNe II-L in the sample in shape, but SN~2013fr is bluer than every other SNe~II-L. It
roughly matches SN~1979C in both shape and colour, and is similarly close in colour to SN~1998S. After about day 50 the points for SN~2013fr
become too noisy to be certain, but if one takes the midpoint value from day 50--70, it remains very similar to SN~1979C and SN~1998S. 

The spectra of SN~2013fr compared with several of the SNe~II-L and SN~2013ej are plotted in Figure \ref{fig:frcompspecIIL}. The narrow H$\alpha$ present in the first spectrum
is similar to that in SN~2013fs. As discussed in Section \ref{res:frevol}, this feature is brightest in the first epoch, has a high EW, and the spectra do not show any other
H\,\textsc{ii} region features. In particular, the EW is $\sim 6$ times larger than at any of the later epochs showing narrow H\,\textsc{ii} region emission. Thus, the narrow
H$\alpha$ emission on day 4 is a potential indicator of brief CSM interaction at early phases.

SN~2013fr has variation in its P Cygni absorption speed at early times, which is atypical in SNe~II-L. Of the SNe~II-L with a P Cygni absorption feature, most have H$\alpha$
velocities $\sim 10,000$ km~s$^{-1}$, with SN~2009kr having the lowest at $v_{{\rm H}\alpha} \approx 7000$ km~s$^{-1}$. SN~2013fr has a P Cygni absorption speed which never
exceeds 5200 km~s$^{-1}$, with FWHM $\approx 2200$~km~s$^{-1}$ after day 7, when the P Cygni absorption is resolved. The spectra of SN~2013fr best match those of SN~2009kr at
similar epochs, though SN~2013fr is $\sim 1.5$—2 mag brighter than SN~2009kr at all epochs. More atypical still is the variation in the P Cygni absorption minimum of
SN~2013fr; it changes by at most $\sim 300$~km~s$^{-1}$. This variation is dramatically smaller than for most of the SNe~II-L which show changes in their P Cygni absorption
minima. Again, SN~2009kr provides the best match, as it is the only other object that shows a similar variation in its P Cygni absorption minimum of $\sim 500$ km~s$^{-1}$.
These small variations, if significant, may provide additional evidence for CSM interaction in SN~2013fr and SN~2009kr.

The SEDs show an IR excess, providing evidence for dense CSM in SN~2013fr, independent of the early narrow emission lines in the day 4 spectrum. The SEDs for days
10-12, 42/43, and 75 are plotted in Figure \ref{fig:frSED}. The second and third epochs show a near-IR excess that cannot be fit by a single temperature. By day 75 the
contrast is substantial; the {\it H}-band flux is comparable to that in the {\it i} and {\it z} bands. This infrared excess can be matched by a two-temperature blackbody,
with $T_{1} = 6200$~K (6300~K) for the SN photosphere and $T_{2} = 1200$~K (1500~K) for the dust in the second (third) SED. Owing to line blanketing after day 28, the {\it B}
and {\it V} points fall below the blackbody curve in the second two epochs. There may, of course, also be cooler dust that is not constrained by these wavelengths.

\begin{table}
  \caption{Evaporation radii for silicate (Si) and graphite (Gr) dust grains of size $a$, given the lower limit $L_{\rm peak} \approx 7.92 \times 10^{42}$~erg~s$^{-1}$, at both
    early times and later.}
  \label{tab:evap}
  \begin{tabular}{cccc}\hline\hline
    $a$      & $T_{\rm eff}$ & $R_{\rm evap}$    & $R_{\rm evap}$ \\
    $\mu$m & K        & $10^{17}$ cm & $10^{17}$ cm \\
           &         &Silicate    &Graphite \\
    \hline
    0.001  &10,000    &2.31        &1.27  \\
    1.0    &10,000    &1.16        &0.303 \\
    \hline
    0.001  &6000     &0.765    &0.765 \\
    1.0    &6000     &0.641    &0.312 \\
    \hline
  \end{tabular}
\end{table}

Following \citet{Fox2011} and \citet{Fransson2013}, we can calculate the evaporation radius for silicate and graphite dust grains, assuming evaporation temperatures of
1500~K (1900~K) for silicates (graphites), respectively. Re-arranging Equation 24 from \citet{Draine1979},

\begin{equation}
  \centering
  R_{\rm evap} = (\frac{L_{\rm peak}}{16 \pi \sigma T^{4}_{\rm evap} \langle Q \rangle})^{\frac{1}{2}},
\end{equation}
\noindent
where $\langle Q \rangle$ is the Planck-averaged dust emissivity, and $T_{\rm evap}$ is the vaporisation temperature of the dust grains. Owing to the later discovery of
SN~2013fr, we can only put an upper limit on $L_{\rm peak}$, since the SN was already fading during the first photometric epochs. To get a lower limit on $L_{\rm peak}$, we use
Equation 7 from \citet{Lyman2013} to obtain the $B$-band bolometric correction, and convert this to a luminosity.

The results are given in Table \ref{tab:evap}. Given a typial shock velocity of 10000-15000~km~s$^{-1}$, the evaporation radii for both silicate and graphite grains of sizes
between 1~nm and 1~$\mu$m lie well above the shock radius ($R_{\rm shock} \approx v_{\rm shock}[t_{\rm disc} + t_{\rm delay}]$, where $t_{\rm delay}$ is the time between explosion and
discovery), assuming the explosion date was within 20 days of discovery.

Such dust could have been formed in situ after the SN explosion, or could have already been present in the CSM prior to the explosion, heated by SN light or by the forward
shock. In the latter case, the dust temperature is correlated to the gas density \citep{Draine1979}, and could be used to probe the mass-loss history of the progenitor. Such
an IR echo results from dust absorption of optical light and subsequent re-emission at IR wavelengths. SNe~IIn and Ibn commonly show IR echoes at late times when the dust is
heated by the radiation from ongoing shock interaction \citep[e.g.,][]{Smith2009c,Fox2010,Mattila2008}, and IR echoes have been detected for decades
after explosion in SN~1980K \citep{Sugerman2012}. The inferred evaporation radius of SN~2013fr renders shock evaporation of dust at these early times unlikely, if the P Cygni
absorption is representative of the velocity of the forward shock. The determined dust temperatures are (especially around day 43) close to the evaporation temperature for
silicate grains, so this dust could be primarily graphite.

Alternatively, if the dust does lie closer, within the ejecta, then the observed echo could still be caused by collisional evaporation of newly formed dust inside the ejecta
by the reverse shock. The presence of an IR excess in SN~2013fr at such early times suggests that CSM is important in at least a subset of SNe~II-L.

In principle, an IR excess could arise from newly formed dust, or an IR echo from pre-existing dust, radiatively heated by SN light or CSM interaction. Such an IR echo
results from dust absorption of optical light and subsequent re-emission at IR wavelengths. SNe IIn and Ibn commonly show IR echoes at late times when the dust is heated by
the radiation from ongoing shock interaction \citep[e.g.,][]{Smith2009c,Fox2010,Mattila2008}, and IR echoes have been detected for decades after
explosion in SN~1980K \citep{Sugerman2012}. The dust temperature in this case is correlated to the gas density \citep{Draine1979}, and can be used to probe the mass-loss
history of the progenitor. Given the dust temperatures in the SEDs and the luminosity of the SN on those epochs, the dust (and thus CSM from pre-SN mass-loss) is located near
the evaporation radius, from 0.5-1.2$\times$10$^{17}$~cm. This separation from the SN ejecta renders the creation of new dust at these epochs unlikely. If we assume a typical
wind speed of 10 km~s$^{-1}$ for this CSM, this mass-loss phase occurred between 16000 and 38000 years before core-collapse.

\section{Conclusions \& Future Prospects}\label{conclusion}
We have presented an analysis of photometry and spectra of SN~2013fs and SN~2013fr. Owing to the good spectroscopic coverage in the first
90~days and the very early discovery time of SN~2013fs, we confirmed the presence of a dense but relatively compact CSM around the progenitor
based on light curve fits and hydrodynamical models by \citet{Morozova2016}, \citet{Takashi2017}, and \citet{Dessart2017}. The
He\,\textsc{ii} features in the day 1-2 spectra, along with the lack of strong CSM interaction after the beginning of the plateau phase point
to a confined CSM shell in the immediate vicinity of the progenitor requiring short-duration enhanced or eruptive mass loss beginning shortly
before core collapse. SN~2013fs is similar to PTF11iqb and other young SNe having WR-like spectral lines, but it is a clear SN~II-P,
providing a demonstration that substantial mass-loss processes in very last stages of massive stellar evolution occur in more than just the
strongly interacting SNe~IIn, and that there is a continuum in late-stage mass-loss rates.

The rise time to maximum brightness of SN~2013fr was undetermined owing to its later discovery relative to the explosion. The first epoch
shows narrow H$\alpha$ emission possibly characteristic of CSM interaction, but this fades quickly. The lack of intrinsic narrow H$\alpha$
emission after the first epoch, combined with the linearly declining light curve of SN~2013fr, show that it is better classified as a
SN~II-L. The strong line blanketing after day 28 puts limits on the strength of later-time CSM interaction, and the unusual shape and low
absorption minima of its P Cygni feature, almost identical to those seen in SN~2009kr, demonstrate that it was a low-velocity, possibly
low-energy explosion. The IR excess that develops later is an indicator of warm dust, and a tracer of the progenitor's mass-loss history.
This feature suggests that dust may be an important factor in the later evolution of SNe~II-L, which are not often followed
more $\gtrsim 100$~days post-explosion.

Short duration eruptive mass-loss has long been investigated for very massive stars due to their proximity to the Eddington limit, owing to
observational evidence for variability and dense CSM \citep{Smith2014} in such stars. Previous work by \citet{Yoon2010} and \citet{Heger1997}
showed that RSGs may also show episodes of enhanced mass-loss, which is typically not included in stellar models. Observations of galactic
RSGs in the last 5-10 years suggest an increase in mass-loss toward the end of stellar evolution along the RSG branch
(\citealp{Davies2008}, \citealp{Smith2014}, \citealp{Beasor2016}).

Both SN~2013fs and SN~2013fr demonstrate the importance of short-lived CSM interaction in classifying SNe. SN~2013fs, in particular,
demonstrates that late-stage enhanced or eruptive mass loss and intrinsic line polarization are not limited to LBVs, and also occur in RSGs.
Had SN~2013fs been discovered a week or more later than it was, the spectroscopic signatures of its close CSM would not have been detected at
all. The same may be true of SN~2013fr, if the narrow emission in the first spectrum is intrinsic SN emission. More precisely determining the
frequency of abrupt late-stage mass loss in the RSG progenitors of SNe~II-P and SNe~II-L will depend on increasing the fraction of these SNe
discovered very soon after explosion. The rapid development of an IR excess in SN~2013fr suggests a CSM dust shell, the result of strong mass
loss over longer timescales than the almost immediate pre-SN mass loss seen in SN~2013fs. Assessing the importance of late-time dust in
SNe~II-L  will require a greater sample of such SNe to be monitored at IR wavelengths for more than 100 days post-explosion.

%% Acknowledgements
\section*{Acknowledgements}
\scriptsize

Observations using Steward Observatory facilities were obtained as part of the observing program AZTEC: Arizona Transient Exploration and Characterization, which receives
support from NSF grant AST-1515559.  This paper includes data obtained by the Supernova Spectropolarimetry Project, supported by the National Science Foundation through grant
AST-1210599. The SN research of AVF's group at U.C. Berkeley is supported by Gary \& Cynthia Bengier, the Richard \& Rhoda Goldman Fund, the Christopher R. Redlich Fund, the
TABASGO Foundation, and NSF grant AST-1211916. KAIT and its ongoing operation were made possible by donations from Sun Microsystems, Inc., the Hewlett-Packard Company,
AutoScope Corporation, Lick Observatory, the NSF, the University of California, the Sylvia \& Jim Katzman Foundation, and the TABASGO Foundation. The Kast spectrograph at Lick
Observatory resulted from a generous donation from Bill and Marina Kast. Research at Lick Observatory is partially supported by a generous gift from Google.

Some observations reported here were obtained at the MMT Observatory, a joint facility of the University of Arizona and the Smithsonian Institution. These results made use of
Lowell Observatory's Discovery Channel Telescope. Lowell operates the DCT in partnership with Boston University, Northern Arizona University, the University of Maryland, and
the University of Toledo. Partial support of the DCT was provided by Discovery Communications. LMI was built by Lowell Observatory using funds from NSF grant AST-1005313. We
made use of {\it Swift}/UVOT data reduced by P. J. Brown and released in the {\it Swift} Optical/Ultraviolet Supernova Archive (SOUSA). SOUSA is supported by NASA's
Astrophysics Data Analysis Program through grant NNX13AF35G. This work is based (in part) on observations collected at the European Organisation for Astronomical Research in
the Southern Hemisphere, Chile as part of PESSTO (the Public ESO Spectroscopic Survey for Transient Objects Survey) ESO programs 188.D-3003 and 191.D-0935. Several of the
spectra were retrieved from WiSEREP, the Weizmann interactive Supernova data REPository (http://wiserep.weizmann.ac.il). We are grateful to Isaac Shivvers and Jeff
Silverman for help with one of the Lick/Kast observations and reductions. We also thank U.C. Berkeley undergraduate students/visitors Andrew Bigley, Kevin Hayakawa, Heechan
Yuk, Minkyu Kim, Kiera Fuller, Philip Lu, James Bradley, Haejung Kim, Chadwick Casper, Gary Li, Samantha Stegman, Kyle Blanchard, Erin Leonard, Jenifer Gross, Xianggao Wang,
Stephen Taylor, and Sahana Kumar for their effort in taking Lick/Nickel data.

We are grateful to the dedicated staffs of the observatories where data for this paper were obtained, Ryan Hofmann for his assistance with the IRAF apphot and daophot routines
used in reducing the DCT/Kuiper data, and Nancy Elias-Rosa for kindly providing comparison spectra of SN~2009kr.

%***Chris: In the bibliography, you need to list the first
% three authors of ``et al.'' papers, and then put ``et al.''

%% Bibliography

\begin{table*}
  \caption{KAIT {\it BVRI} photometry of SN~2013fs.}
  \label{tab:lickfs}
  \centering
  \begin{tabular}{ccccccccc}
    \hline
    $t$ & $B$ &$\sigma_{B}$ & $V$ &$\sigma_{V}$ & $R$ &$\sigma_{R}$ & $I$ &$\sigma_{I}$ \\
    (MJD) &(mag) &(mag) &(mag) &(mag) &(mag) &(mag) &(mag) &(mag) \\
    \hline
    %% 2013fs Lick data
    56575.3  &16.19 &0.1  &16.04 &0.04 &16.03 &0.03 &15.92 &0.04 \\
    56576.2  &16.04 &0.09 &16.02 &0.04 &15.98 &0.03 &15.85 &0.03 \\
    56577.2  &16.11 &0.1  &16.02 &0.04 &15.89 &0.03 &15.78 &0.04 \\
    56578.2  &16.18 &0.14 &15.97 &0.05 &15.87 &0.03 &15.69 &0.04 \\
    56579.2  &16.19 &0.12 &16.00 &0.05 &15.84 &0.03 &15.71 &0.03 \\
    56585.3  &16.44 &0.16 &16.21 &0.06 &15.97 &0.04 &15.79 &0.04 \\
    56586.3  &16.62 &0.16 &16.26 &0.05 &16.06 &0.03 &15.82 &0.04 \\
    56587.2  &16.54 &0.10 &16.36 &0.05 &16.06 &0.04 &15.86 &0.05 \\
    56588.2  &16.51 &0.11 &16.37 &0.05 &16.12 &0.03 &15.90 &0.04 \\
    56589.2  &16.65 &0.09 &16.42 &0.04 &16.11 &0.03 &15.88 &0.04 \\
    56590.2  &16.67 &0.10 &16.43 &0.04 &16.13 &0.03 &15.95 &0.04 \\
    56591.2  &16.77 &0.07 &16.46 &0.04 &16.19 &0.03 &16.01 &0.04 \\
    56592.2  &16.72 &0.07 &16.48 &0.04 &16.16 &0.03 &16.02 &0.04 \\
    56596.2  &16.76 &0.09 &16.53 &0.04 &16.20 &0.03 &16.03 &0.03 \\
    56597.2  &16.94 &0.09 &16.51 &0.03 &16.22 &0.02 &16.06 &0.04 \\
    56598.2  &17.05 &0.07 &16.54 &0.03 &16.24 &0.03 &16.06 &0.04 \\
    56599.2  &17.03 &0.14 &16.52 &0.06 &16.15 &0.05 &-     &- \\
    56600.2  &17.04 &0.21 &16.60 &0.07 &16.18 &0.05 &16.04 &0.07 \\
    56601.2  &17.12 &0.09 &16.55 &0.04 &16.23 &0.02 &16.05 &0.03 \\
    56604.2  &17.26 &0.07 &16.58 &0.03 &16.29 &0.03 &16.10 &0.04 \\
    56605.2  &17.35 &0.09 &16.63 &0.04 &16.31 &0.03 &16.13 &0.04 \\
    56607.2  &17.34 &0.14 &16.72 &0.07 &16.34 &0.11 &16.04 &0.07 \\
    56611.2  &17.56 &0.19 &16.65 &0.08 &16.24 &0.03 &16.02 &0.04 \\
    56614.2  &17.54 &0.38 &16.63 &0.15 &16.34 &0.08 &16.18 &0.10 \\
    56619.2  &17.62 &0.13 &16.66 &0.05 &16.30 &0.03 &15.97 &0.05 \\
    56621.2  &17.65 &0.1  &16.76 &0.04 &16.36 &0.03 &16.10 &0.04 \\
    %% end
    \hline
  \end{tabular}
\end{table*}

\begin{table}
  \caption{Unfiltered KAIT photometry of SN~2013fs. Photometrically, the clear filter is similar to Johnson-Cousins R.}
  \label{tab:clearfs}
  \begin{tabular}{ccc}
    \hline
    $t$ & $C$ & $\sigma_{C}$ \\
    (MJD) &(mag) &(mag) \\
    \hline
    %% Clear data start
    56572.23 &16.26 &0.08 \\
    56574.23 &15.88 &0.03 \\
    56575.26 &15.80 &0.04 \\
    56575.28 &15.85 &0.02 \\
    56576.24 &15.79 &0.02 \\
    56577.24 &15.72 &0.03 \\
    56577.28 &15.75 &0.02 \\
    56578.21 &15.74 &0.03 \\
    56579.24 &15.68 &0.02 \\
    56585.25 &15.89 &0.03 \\
    56586.27 &15.92 &0.03 \\
    56587.23 &15.93 &0.03 \\
    56588.23 &15.98 &0.02 \\
    56589.22 &16.04 &0.03 \\
    56590.23 &16.05 &0.04 \\
    56592.22 &16.15 &0.03 \\
    56596.23 &16.15 &0.03 \\
    56597.21 &16.18 &0.02 \\
    56598.18 &16.16 &0.03 \\
    56599.18 &16.06 &0.03 \\
    56600.18 &16.17 &0.02 \\
    56601.18 &16.20 &0.02 \\
    56603.21 &16.31 &0.10 \\
    56604.18 &16.23 &0.02 \\
    56605.20 &16.24 &0.10 \\
    56607.17 &16.28 &0.03 \\
    56611.22 &16.27 &0.03 \\
    56612.18 &16.37 &0.06 \\
    56614.18 &16.34 &0.04 \\
    56615.18 &16.57 &0.20 \\
    56618.14 &16.28 &0.03 \\
    56619.17 &16.19 &0.03 \\
    56620.15 &16.39 &0.03 \\
    56621.17 &16.39 &0.03 \\
    56622.13 &16.39 &0.04 \\
    56623.16 &16.54 &0.19 \\
    56624.14 &16.39 &0.05 \\
    56625.17 &16.43 &0.04 \\
    56626.13 &16.41 &0.04 \\
    56627.13 &16.42 &0.04 \\
    56628.14 &16.39 &0.03 \\
    56629.14 &16.37 &0.04 \\
    56630.14 &16.42 &0.03 \\
    56631.11 &16.45 &0.04 \\
    56632.11 &16.46 &0.04 \\
    56638.14 &16.54 &0.05 \\
    56640.12 &16.61 &0.04 \\
    56642.11 &16.59 &0.06 \\
    56643.13 &16.65 &0.07 \\
    56647.14 &16.69 &0.07 \\
    56648.13 &16.79 &0.04 \\
    56650.12 &16.92 &0.05 \\
    56652.12 &17.11 &0.04 \\
    56653.11 &17.17 &0.06 \\
    56656.13 &17.46 &0.06 \\
    56658.08 &17.58 &0.10 \\
    56660.08 &17.68 &0.16 \\
    56662.10 &17.97 &0.13 \\
    56665.10 &17.94 &0.13 \\
    56674.12 &18.04 &0.07 \\
    56681.12 &17.90 &0.13 \\
    %% end
    \hline
  \end{tabular}
\end{table}

\begin{table*}
  \caption{Lick photometry of SN~2013fr; data come from both the KAIT and Nickel telescopes, and are listed together here. Distinct observations from within 24~hr
    of each other indicate a day on which SN 2013fr was observed with both KAIT and Nickel.}
  \label{tab:lickfr}
  \centering
  \begin{tabular}{ccccccccc}
    \hline
    Time & $B$ & $\sigma_{B}$ & $V$ &$\sigma_{V}$ & $R$ &$\sigma_{R}$ & $I$ &$\sigma_{I}$ \\
    MJD &(mag) &(mag) &(mag) &(mag) &(mag) &(mag) &(mag) &(mag) \\
    \hline
    %% 2013fr Lick data
    56573.4 &17.02 &0.02 &16.93 &0.01 &16.65 &0.01 &16.49 &0.02 \\
    56575.3 &17.25 &0.01 &16.97 &0.01 &16.70 &0.01 &16.51 &0.01 \\
    56581.5 &17.50 &0.01 &17.13 &0.01 &16.79 &0.01 &16.57 &0.01 \\
    56585.5 &17.63 &0.03 &17.18 &0.02 &16.85 &0.02 &16.65 &0.03 \\
    56587.5 &17.77 &0.28 &17.25 &0.19 &16.95 &0.13 &16.74 &0.21 \\
    56589.5 &17.95 &0.02 &17.31 &0.01 &16.95 &0.02 &16.79 &0.02 \\
    56589.5 &17.79 &0.11 &16.85 &0.05 &16.89 &0.05 &16.63 &0.07 \\
    56591.5 &17.91 &0.10 &17.47 &0.06 &16.98 &0.05 &16.76 &0.05 \\
    56597.4 &18.15 &0.08 &17.41 &0.05 &17.08 &0.04 &16.71 &0.04 \\
    56602.4 &18.47 &0.12 &17.61 &0.05 &17.13 &0.04 &16.91 &0.05 \\
    56604.4 &18.69 &0.16 &17.76 &0.06 &17.14 &0.04 &16.87 &0.05 \\
    56605.5 &18.49 &0.02 &17.66 &0.01 &17.17 &0.02 &16.87 &0.02 \\
    56609.5 &18.94 &0.03 &17.78 &0.02 &17.28 &0.02 &16.94 &0.02 \\
    56610.3 &18.68 &0.21 &17.98 &0.10 &17.46 &0.07 &16.99 &0.06 \\
    56618.3 &19.10 &0.36 &18.29 &0.16 &17.89 &0.09 &17.49 &0.12 \\
    56621.4 &19.75 &0.37 &18.35 &0.08 &17.76 &0.06 &17.50 &0.07 \\
    56622.3 &19.94 &0.12 &18.28 &0.03 &17.67 &0.02 &17.36 &0.02 \\
    %% end
    \hline
  \end{tabular}
\end{table*}

\begin{table}
  \caption{Later-time limiting magnitudes of SN~2013fs, obtained by the 61~inch Kuiper telescope, as discussed in Section \ref{obs:Kuiper}. Late-time limits were also obtained
    of SN~2013fr on day 139, and these limits are found in Section \ref{obs:DCT}.}.
  \label{tab:kuiper}
  \begin{tabular}{ccccccccccccccc}
    \hline
    Time & $B$ & $V$ & $R$ & $I$ \\
    MJD &(mag) &(mag) &(mag) &(mag) \\
    \hline
    %% Kuiper data
    56960 &20.1 &20.6 &21.2  &20.4 \\
    56990 &20.3 &20.7 &21.5  &20.6  \\
    57010 &19.7 &20.1 &21.0  &19.2  \\
    %% End Kuiper
    \hline
  \end{tabular}
\end{table}

\begin{table*}
  \caption{RATIR $rizYJH$ photometry of SN~2013fr. All magnitudes are reported on the AB system.}
  \label{tab:RATIR}
  \centering
  \begin{tabular}{ccccccccccccc}
    \hline
    Time & $r$ &$\sigma_{r}$ & $i$ &$\sigma_{i}$ & $z$ &$\sigma_{z}$ & $Y$ &$\sigma_{Y}$ & $J$ &$\sigma_{J}$ & $H$ &$\sigma_{H}$ \\
    days &(mag) &(mag) &(mag) &(mag) &(mag) &(mag) &(mag) &(mag) &(mag) &(mag) &(mag) &(mag) \\
    \hline
    %% RATIR data
    56570.0  &-     &-    &-     &-    &16.78 &0.01 &16.65 &0.06 &16.95 &0.06 &-     &-    \\
    56570.3  &-     &-    &16.72 &0.01 &-     &-    &-     &-    &-     &-    &-     &-    \\
    56572.0  &-     &-    &-     &-    &16.83 &0.04 &16.64 &0.06 &-     &-    &16.90 &0.25 \\
    56572.3  &-     &-    &16.76 &0.01 &-     &-    &-     &-    &-     &-    &-     &-    \\
    56573.0  &-     &-    &-     &-    &16.77 &0.02 &16.68 &0.05 &16.92 &0.06 &-     &-    \\
    56573.3  &16.82 &0.01 &16.80 &0.01 &-     &-    &-     &-    &-     &-    &-     &-    \\
    56576.0  &-     &-    &-     &-    &16.88 &0.04 &16.68 &0.05 &17.02 &0.06 &16.88 &0.22 \\
    56576.3  &16.89 &0.01 &16.84 &0.01 &-     &-    &-     &-    &-     &-    &-     &-    \\
    56579.0  &-     &-    &-     &-    &16.89 &0.03 &16.69 &0.07 &17.00 &0.06 &17.02 &0.07 \\
    56579.3  &16.97 &0.02 &16.88 &0.03 &-     &-    &-     &-    &-     &-    &-     &-    \\
    56581.0  &-     &-    &-     &-    &16.87 &0.01 &16.71 &0.17 &16.99 &0.06 &-     &-    \\
    56581.2  &16.99 &0.02 &16.92 &0.02 &-     &-    &-     &-    &-     &-    &-     &-    \\
    56584.0  &-     &-    &-     &-    &16.89 &0.02 &16.57 &0.20 &16.94 &0.06 &-     &-    \\
    56584.3  &17.00 &0.02 &16.92 &0.01 &-     &-    &-     &-    &-     &-    &-     &-    \\
    56588.0  &-     &-    &-     &-    &16.93 &0.02 &16.64 &0.19 &17.04 &0.06 &16.96 &0.05 \\
    56588.2  &17.08 &0.01 &16.97 &0.01 &-     &-    &-     &-    &-     &-    &-     &-    \\
    56590.0  &-     &-    &-     &-    &16.97 &0.02 &-     &-    &17.02 &0.06 &-     &-    \\
    56597.0  &-     &-    &-     &-    &17.02 &0.02 &16.85 &0.07 &16.97 &0.06 &16.99 &0.10 \\
    56597.5  &17.21 &0.03 &17.06 &0.03 &-     &-    &-     &-    &-     &-    &-     &-    \\
    56603.0  &-     &-    &-     &-    &17.15 &0.03 &16.95 &0.06 &17.08 &0.05 &17.07 &0.09 \\
    56603.4  &17.34 &0.02 &17.16 &0.02 &-     &-    &-     &-    &-     &-    &-     &-    \\
    56607.0  &-     &-    &-     &-    &17.21 &0.04 &16.93 &0.06 &16.98 &0.07 &16.96 &0.07 \\
    56607.2  &17.42 &0.02 &17.24 &0.03 &-     &-    &-     &-    &-     &-    &-     &-    \\
    56612.0  &-     &-    &-     &-    &17.37 &0.02 &17.18 &0.07 &17.21 &0.05 &17.18 &0.10 \\
    56612.2  &17.58 &0.03 &17.38 &0.02 &-     &-    &-     &-    &-     &-    &-     &-    \\
    56613.0  &-     &-    &-     &-    &17.36 &0.06 &16.94 &0.05 &-     &-    &-     &-    \\
    56613.2  &17.62 &0.01 &17.27 &0.02 &-     &-    &-     &-    &-     &-    &-     &-    \\
    56627.0  &-     &-    &-     &-    &17.79 &0.01 &17.57 &0.06 &17.48 &0.04 &17.28 &0.07 \\
    56627.3  &18.17 &0.04 &17.91 &0.04 &-     &-    &-     &-    &-     &-    &-     &-    \\
    56630.0  &-     &-    &-     &-    &17.94 &0.04 &17.54 &0.09 &17.87 &0.05 &17.31 &0.10 \\
    56630.1  &18.30 &0.01 &17.98 &0.01 &-     &-    &-     &-    &-     &-    &-     &-    \\
    56633.0  &18.42 &0.02 &18.18 &0.02 &18.02 &0.03 &17.54 &0.07 &17.42 &0.05 &-     &-    \\
    56639.0  &18.99 &0.01 &18.71 &0.02 &18.53 &0.04 &18.14 &0.07 &18.19 &0.06 &17.67 &0.07 \\
    %% end RATIR
    \hline
  \end{tabular}
\end{table*}

\begin{table*}
  \caption{{\it Swift} UVOT photometry, taken up to day 23.6 after explosion. These data were obtained from SOUSA and reduced by Peter Brown, using a template image taken
  on 2016 July 14.}
  \label{tab:uvot}
  \centering
  \begin{tabular}{ccccccccccccc}\hline\hline
    Time & $UVW2$ &$\sigma_{UVW2}$ & $UVM2$ &$\sigma_{UVM2}$ & $UVW1$ &$\sigma_{UVW1}$ & $U$ &$\sigma_{U}$ & $B$ &$\sigma_{B}$ & $V$ &$\sigma_{V}$ \\
    MJD &(mag) &(mag) &(mag) &(mag) &(mag) &(mag) &(mag) &(mag) &(mag) &(mag) &(mag) &(mag) \\
    \hline
    %% UVOT data:
    56572   &14.094 &0.044 &14.327 &0.057 &14.553 &0.055 &-      &-     &-      &-     &-      &- \\
    56573.8 &14.123 &0.045 &14.142 &0.051 &14.304 &0.049 &-      &-     &-      &-     &-      &- \\
    56575.0 &14.414 &0.049 &14.356 &0.059 &14.449 &0.05  &14.763 &0.048 &16.072 &0.058 &16.083 &0.073 \\
    56575.9 &14.817 &0.055 &14.776 &0.064 &14.661 &0.053 &14.829 &0.049 &16.093 &0.058 &16.04  &0.072 \\
    56576.4 &14.791 &0.054 &14.742 &0.064 &14.660 &0.055 &14.836 &0.05  &16.028 &0.059 &15.966 &0.065 \\
    56577.3 &15.034 &0.062 &14.950 &0.067 &14.865 &0.063 &14.905 &0.052 &16.014 &0.06  &15.964 &0.063 \\
    56578.7 &15.49  &0.074 &15.254 &0.071 &15.139 &0.063 &15.049 &0.055 &16.081 &0.061 &15.984 &0.063 \\
    56580.1 &15.773 &0.078 &15.599 &0.079 &15.43  &0.069 &15.132 &0.058 &16.114 &0.064 &16.02  &0.078 \\
    56583.6 &16.563 &0.088 &16.492 &0.098 &16.07  &0.076 &15.501 &0.067 &16.31  &0.065 &16.094 &0.076 \\
    56584.4 &16.824 &0.091 &16.732 &0.095 &16.192 &0.075 &15.562 &0.065 &16.428 &0.064 &16.279 &0.075 \\
    56585.5 &16.94  &0.094 &16.993 &0.106 &16.375 &0.078 &15.671 &0.066 &16.452 &0.064 &16.158 &0.073 \\
    56586.4 &17.155 &0.099 &17.142 &0.11  &16.552 &0.08  &15.776 &0.066 &16.514 &0.064 &16.294 &0.074 \\
    56587.6 &17.35  &0.111 &17.52  &0.142 &16.617 &0.087 &15.902 &0.07  &16.526 &0.067 &16.322 &0.081 \\
    56588.5 &17.616 &0.122 &17.492 &0.142 &16.874 &0.093 &15.916 &0.07  &16.562 &0.066 &16.462 &0.083 \\
    56589.4 &17.731 &0.127 &17.896 &0.169 &16.971 &0.095 &16.10  &0.072 &16.619 &0.067 &16.409 &0.081 \\
    56590.5 &17.95  &0.139 &18.149 &0.195 &17.172 &0.1   &16.143 &0.071 &16.587 &0.065 &16.443 &0.079 \\
    56591.4 &-      &-     &-      &-     &17.742 &0.145 &16.242 &0.085 &16.761 &0.17  &-      &- \\
    56592.9 &18.7   &0.219 &18.709 &0.294 &17.611 &0.126 &16.564 &0.08  &16.742 &0.068 &16.52  &0.083 \\
    56593.1 &18.779 &0.232 &-      &-     &17.738 &0.135 &16.608 &0.082 &16.764 &0.069 &16.435 &0.081 \\
    56594.7 &18.946 &0.259 &-      &-     &18.146 &0.169 &16.723 &0.084 &16.866 &0.07  &16.572 &0.085 \\
    56595.6 &19.209 &0.314 &-      &-     &17.972 &0.153 &16.796 &0.087 &16.927 &0.071 &16.583 &0.086 \\
    %% End UVOT
    \hline
  \end{tabular}
\end{table*}

\begin{table*}
  \caption{All spectra of SN~2013fs and SN~2013fr obtained by Arizona or Berkeley observing facilities. All other spectra were retrieved from WISeREP or other online
  repositories of spectra. Days are given post-discovery: 2013 Oct. 7 for SN~2013fs, 2013 Sep. 28 for SN~2013fr.}
  \label{tab:spectra}
  \begin{tabular}{ccccc}\hline\hline
    Days  &Telescope/Instrument &$\Delta\lambda$ (\AA) &Res. (\AA) &SN \\
    4     &Kuiper/SPol          &3850--7800  &3.95  &2013fs \\
    5     &Kuiper/SPol          &3900--7900  &3.95  &2013fs \\
    6     &Kuiper/SPol          &3850--7800  &3.95  &2013fs \\
    21    &MMT/SPol             &4000--7050  &2.47  &2013fs \\
    23    &MMT/SPol             &4000--7050  &2.47  &2013fs \\
    52    &Bok/SPol             &3850--7800  &3.95  &2013fs \\
    57    &Bok/SPol             &3850--7800  &3.95  &2013fs \\
    86    &Bok/SPol             &3860--7800  &3.95  &2013fs \\
    87    &Bok/SPol             &3860--7800  &3.95  &2013fs \\
    87    &MMT/BlueChannel      &5690--6850  &0.43  &2013fs \\
    \hline
    7     &Lick/Kast            &3350--10,250 &1.96  &2013fr \\
    12    &Lick/Kast            &3350--10,250 &1.96  &2013fr \\
    28    &Lick/Kast            &3350--9800  &1.96  &2013fr \\
    35    &Lick/Kast            &3350--10,250 &1.96  &2013fr \\
    41    &Lick/Kast            &3380--10,380 &1.96  &2013fr \\
    46    &Magellan/IMACS       &5280--9460  &1.46  &2013fr \\
    69    &Lick/Kast            &3370--10,380 &1.96  &2013fr \\
    \hline
  \end{tabular}
\end{table*}

\afterpage{\clearpage}

\newpage

\afterpage{\clearpage}

\newpage

%%%%%%%%%%%%%%%%%%%%%%%%%%%%%%%%%

\bsp
\label{lastpage}

\begin{thebibliography}{99}
\bibitem[\protect\citeauthoryear{Anderson et al.}{2014}]{Anderson2014}
  Anderson~J.~P. et al., 2014, ApJ, 786, 67
\bibitem[\protect\citeauthoryear{Arcavi et al.}{2012}]{Arcavi2012}
  Arcavi~I. et al., 2012, ApJ, 756, L30
\bibitem[\protect\citeauthoryear{Barbon et al.}{1982}]{Barbon1982}
  Barbon~R. et al., 1982, A\&A, 116, 43
\bibitem[\protect\citeauthoryear{Beasor \& Davies}{2016}]{Beasor2016}
  Beasor~E.~R., Davies~B., MNRAS, 463, 1269
\bibitem[\protect\citeauthoryear{Benetti et al.}{1994}]{Benetti1994}
  Benetti~S. et al., 1994, A\&A, 285, L13
\bibitem[\protect\citeauthoryear{Bilinski et al.}{2015}]{Bil2015}
  Bilinski~C. et al., 2015, MNRAS, 450, 246
\bibitem[{{Blondin} \& {Tonry}(2007)}]{Blondin07} 
  Blondin~S., Tonry~J.~L. 2007, ApJ, 666, 1024
\bibitem[\protect\citeauthoryear{Breeveld et al.}{2011}]{Bre2011}
  Breeveld~A.~A. et al., 2011, in AIP Conf. Ser. 1358, eds. J.E. McEnery, J.L. Racusin, \& N. Gehrels, 373
\bibitem[\protect\citeauthoryear{Brown et al.}{2009}]{Brown2009}
  Brown~P. et al., 2009, AJ, 137, 4517
\bibitem[\protect\citeauthoryear{Brown et al.}{2014}]{Brown2014}
  Brown~P. et al., 2014, Ap\&SS, 354, 89
\bibitem[\protect\citeauthoryear{Butler et al.}{2012}]{butler12}
  Butler, N., Klein, C., Fox, O., et al., 2012, Proc. of the SPIE, 8446, 10
\bibitem[\protect\citeauthoryear{Cardelli, Clayton \& Mathis}{1989}]{Cardelli1989}
  Cardelli~J.~A., Clayton~G.~C., Mathis~J.~S., 1989, ApJ, 345, 245
\bibitem[\protect\citeauthoryear{Childress et al.}{2013}]{ATEL}
  Childress~M.~J. et al., 2013, ATEL, 5527, 1
\bibitem[\protect\citeauthoryear{Childress et al.}{2016}]{Child2016}
  Childress~M.~J. et al., 2016, arXiv:1607.08526
\bibitem[\protect\citeauthoryear{Davies et al.}{2008}]{Davies2008}
  Davies~B., Figer~D.~F., Law~C.~J. et al., 2008, ApJ, 676, 1016
\bibitem[\protect\citeauthoryear{Dessart \& Hillier}{2011}]{Dessart2011}
  Dessart~L., Hillier~D.~J., 2011, MNRAS, 415, 3497
\bibitem[\protect\citeauthoryear{Dessart, Hillier, and Audit}{2017}]{Dessart2017}
  Dessart~L., Hillier~D.J., Audit~E., A\&A, 605, A83
\bibitem[\protect\citeauthoryear{Draine \& Salpeter}{1979}]{Draine1979}
  Draine~B.~T., Salpeter~E.~E., 1979, ApJ, 231, 77
\bibitem[\protect\citeauthoryear{Dressler et al.}{2011}]{Dressler2011}
  Dressler~A. et al., 2011, PASP, 123, 901
\bibitem[\protect\citeauthoryear{Elias-Rosa et al.}{2010}]{EliasRosa2010}
  Elias-Rosa~N. et al., 2010, ApJ, 714, L254
\bibitem[\protect\citeauthoryear{Elmhamdi et al.}{2003}]{Elmhamdi2003}
  Elmhamdi~A. et al., 2003, MNRAS 338, 939
\bibitem[\protect\citeauthoryear{Faran et al.}{2014a}]{Faran2014-2}
  Faran~T., et al., 2014a, MNRAS, 442, 844
\bibitem[\protect\citeauthoryear{Faran et al.}{2014b}]{Faran2014}
  Faran~T., et al., 2014b, MNRAS, 445, 554
\bibitem[\protect\citeauthoryear{Fassia et al.}{2001}]{Fassia2001}
  Fassia~A. et al., 2001, MNRAS, 325, 907
\bibitem[\protect\citeauthoryear{Fillippenko}{1997}]{Filippenko1997}
  Filippenko~A.~V., 1997, ARA\&A, 35, 309
\bibitem[\protect\citeauthoryear{Filippenko et al.}{2001}]{filippenko2001}
  Filippenko~A.V., Li~W.~D., Treffers~R.~R., Modjaz~M., 2001, in {Small-Telescope Astronomy on Global Scales.}, ed. B.~{Paczy\'{n}ski}, W.~P.
 {Chen}, \& C.~{Lemme} (San Francisco: ASP), 121
\bibitem[\protect\citeauthoryear{Foley et al.}{2003}]{Foley2003}
  Foley~R.~J. et al., 2003, PASP, 115, 1220
\bibitem[\protect\citeauthoryear{Foley et al.}{2007}]{Foley2007}
  Foley~R.~J. et al., 2007, ApJ, 657, L105
\bibitem[\protect\citeauthoryear{Fox et al.}{2010}]{Fox2010}
  Fox~O.~D. et al., 2010, ApJ, 725, 1768
\bibitem[\protect\citeauthoryear{Fox et al.}{2011}]{Fox2011}
  Fox~O.~D. et al., 2011, ApJ, 741, 7
\bibitem[\protect\citeauthoryear{Fox et al.}{2012}]{fox12}
  Fox O. D., Kutyrev A. S., Rapchun D. A., et al., 2012, Proc. of SPIE, 8453, 59
\bibitem[\protect\citeauthoryear{Fransson et al.}{2005}]{Fransson2005}
  Fransson~C. et al., 2005, ApJ, 622, 991
\bibitem[\protect\citeauthoryear{Fransson et al.}{2013}]{Fransson2013}
  Fransson~C. et al., 2013, ApJ, 797, 118
\bibitem[\protect\citeauthoryear{Galbany et al.}{2016}]{Galbany2016}
  Galbany~L. et al., AJ, 151, 33
\bibitem[\protect\citeauthoryear{Gal-Yam et al.}{2014}]{GalYam2014}
  Gal-Yam~A. et al., 2014, Nature, 509, 471
\bibitem[\protect\citeauthoryear{Ganeshalingam et al.}{2010}]{Ganeshalingam2010}
  Ganeshalingam~M., Li~W., Filippenko~A.~V., et al., 2010, ApJS, 190, 418
\bibitem[\protect\citeauthoryear{Groh}{2014}]{Groh2014}
  Groh~Jose~H., 2014, A\&A, 572, L11
\bibitem[\protect\citeauthoryear{Guillochon, Parrent, \& Margutti.}{2016}]{Guillochon2016}
  Guillochon~J., Parrent~J., Margutti~R. 2016, arXiv:1605.01054
\bibitem[\protect\citeauthoryear{Hamuy et al.}{2001}]{Hamuy2001}
  Hamuy~M. et al., 2001, ApJ, 558, 615
\bibitem[\protect\citeauthoryear{Heger et al.}{1997}]{Heger1997}
  Heger~A., Jeannin~L., Langer~N. et al., 1997, A\&A, 327, 224
\bibitem[\protect\citeauthoryear{Hillier \& Miller}{1998}]{Hillier1998}
  Hillier,~D.~J. \& Miller,~D.~L., ApJ, 496, 407
\bibitem[\protect\citeauthoryear{Howerton et al.}{2013}]{CBET3666}
  Howerton~S., Drake~A. et al., 2013, CBET, 3666, 1
\bibitem[\protect\citeauthoryear{Huang et al.}{2015}]{Huang2015}
  Huang~F. et al., 2015, ApJ, 807, 59
\bibitem[\protect\citeauthoryear{Huang et al.}{2016}]{Huang2016}
  Huang~F. et al., 2016, ApJ, 832, 139
\bibitem[\protect\citeauthoryear{Khazov et al.}{2016}]{Khazov2016}
  Khazov~D., et al. 2016, ApJ, 818, 3
\bibitem[\protect\citeauthoryear{Komatsu et al.}{2009}]{Komatsu2009}
  Komatsu~E., et al. 2009, ApJS, 180, 330
\bibitem[\protect\citeauthoryear{Leonard et al.}{2000}]{Leonard2000}
  Leonard~D.~C. et al., 2000, ApJ, 536, 239
\bibitem[\protect\citeauthoryear{Leonard et al.}{2002}]{Leonard2002}
  Leonard~D.~C. et al., 2002, PASP, 114, 35
\bibitem[\protect\citeauthoryear{Lyman, Bersier, and James}{2013}]{Lyman2013}
  Lyman~J.~D., Bersier~D., James~P.~A., 2013, MNRAS, 437, 3848
\bibitem[\protect\citeauthoryear{Mattila et al.}{2008}]{Mattila2008}
  Mattila~S. et al., 2008, MNRAS, 389, 141
\bibitem[\protect\citeauthoryear{Mauerhan et al.}{2013}]{Mauerhan2013}
  Mauerhan~J. et al., 2013, MNRAS, 430, 1801
\bibitem[\protect\citeauthoryear{Mauerhan et al.}{2016}]{Mauerhan2016}
  Mauerhan~J. et al., 2016, arXiv:1611:07930
\bibitem[\protect\citeauthoryear{Miller \& and Stone}{1993}]{Miller1993}
  Miller~J.S., Stone~R.P.S., 1993, Lick Obs. Tech. Rep. 66 (Santa Cruz: Lick Obs.)
\bibitem[\protect\citeauthoryear{Morozova et. al}{2015}]{Morozova2015}
  Morozova~V., Piro~A.L., Renzo~M, et al., 2015, ApJ, 814, 63
\bibitem[\protect\citeauthoryear{Morozova, Piro, \& Valenti}{2017}]{Morozova2016}
  Morozova~V., Piro~A.L., Valenti~S., 2017, ApJ, 838, 28
\bibitem[\protect\citeauthoryear{Nakano et al.}{2013}]{CBET3671}
  Nakano~S. et al., 2013, CBET, 3671, 1
\bibitem[\protect\citeauthoryear{Niemala, Ruiz, \& Phillips}{1985}]{Niemala1985}
  Niemala~V.~S., Ruiz~M.~T., Phillips~M.~M., 1985, ApJ, 289, 52
\bibitem[\protect\citeauthoryear{Pritchard et al.}{2013}]{Pritchard2013}
  Pritchard~T.A. et al. 2013, ApJ, 787, 157
\bibitem[\protect\citeauthoryear{Quataert \& Shiode}{2012}]{Quataert2012}
  Quataert~E., Shiode,~J., 2012, MNRAS, 423, L92
\bibitem[\protect\citeauthoryear{Quimby et al.}{2007}]{Quimby2007}
  Quimby~R.~M. et al., 2007, ApJ, 666, 1093
\bibitem[\protect\citeauthoryear{Roming et al.}{2005}]{Rom2005}
  Roming~P.W.A., Kennedy~T.E., Mason~K.O., et al. 2005, Space Science Reviews, 120, 95
\bibitem[\protect\citeauthoryear{Sanders et al.}{2015}]{Sanders2015}
  Sanders~N.~E. et al., 2015, ApJ, 799, 208
\bibitem[\protect\citeauthoryear{Schlafly \& Finkbeiner}{2011}]{Schlafly2011}
  Schlafly~E., Finkbeiner~D., 2011, ApJ, 737, 103
\bibitem[\protect\citeauthoryear{Schlegel et al.}{1990}]{Schlegel1990}
  Schlegel~E.M., 1990, MNRAS, 244, 269
\bibitem[\protect\citeauthoryear{Schmidt et al.}{1992}]{Schmidt1992}
  Schmidt~G.~D. et al., 1992, AJ, 104, 1563
\bibitem[\protect\citeauthoryear{Shivvers et al.}{2015}]{Shivvers2015}
  Shivvers~I. et al., 2015, ApJ, 806, 213
\bibitem[\protect\citeauthoryear{Smartt et al.}{2002}]{Smartt2002}
  Smartt~S.~J. et al., 2002, ApJ, 565, 1089
\bibitem[\protect\citeauthoryear{Smartt et al.}{2015}]{PESSTO}
  Smartt~S.~J. et al., 2015, A\&A, 579, A40
\bibitem[\protect\citeauthoryear{Smith}{2014}]{Smith2014}
  Smith~N., 2014, ARA\&A, 52, 487
\bibitem[\protect\citeauthoryear{Smith \& Arnett}{2014}]{SmithArnett2014}
  Smith~N., Arnett~W.~D., 2014, ApJ, 785, 82
\bibitem[\protect\citeauthoryear{Smith \& Hartigan}{2006}]{SmithHart2006}
  Smith~N., Hartigan~P., 2006, ApJ, 638, 1045
\bibitem[\protect\citeauthoryear{Smith, Hinkle, \& Ryde}{2009}]{Smith2009}
  Smith~N., Hinkle~K.~H., Ryde~N., 2009, AJ, 137, 3558
\bibitem[\protect\citeauthoryear{Smith \& Owocki}{2006}]{Smith2006}
  Smith~N., Owocki~S.~P., 2006, ApJ, 645, L45
\bibitem[\protect\citeauthoryear{Smith et al.}{2003}]{Smith2003}
  Smith~N. et al., 2003, AJ, 125, 1458
\bibitem[\protect\citeauthoryear{Smith et al.}{2009c}]{Smith2009c}
  Smith~N. et al., 2009c, ApJ, 695, 1334
\bibitem[\protect\citeauthoryear{Smith et al.}{2011a}]{Smith2011}
  Smith~N. et al., 2011a, MNRAS, 412, 1522
\bibitem[\protect\citeauthoryear{Smith et al.}{2015}]{Smith2015}
  Smith~N. et al., 2015, MNRAS, 449, 1876
\bibitem[\protect\citeauthoryear{Sugerman et al.}{2012}]{Sugerman2012}
  Sugerman~B.~E.~K. et al., 2012, ApJ, 749, 170
\bibitem[\protect\citeauthoryear{Takashi et al.}{2017}]{Takashi2017}
  Takashi~M.~J., Yoon~S.-C., Gr{\"a}fener~G., et al., 2017, MNRAS, 469, L108
\bibitem[\protect\citeauthoryear{Valenti et al.}{2015}]{Valenti2015}
  Valenti~S. et al., 2015, MNRAS, 448, 2608
\bibitem[\protect\citeauthoryear{Valenti et al.}{2016}]{Valenti2016}
  Valenti~S. et al., 2016, MNRAS, 459, 3939
\bibitem[\protect\citeauthoryear{Wardle \& Kronberg}{1974}]{Wardle1974}
  Wardle~J.~F.~C., Kronberg~P.~P., 1974, ApJ, 194, 249
\bibitem[\protect\citeauthoryear{Watson et al.}{2012}]{watson12}
  Watson A. M., Richer M. G., Bloom J. S. et al., 2012, Proc. of SPIE, 8444, doi:10.1117/12.926927
\bibitem[\protect\citeauthoryear{Yaron \& Gal-Yam}{2012}]{WiSEREP}
  Yaron~O., Gal-Yam~A., 2012, PASP, 116, 326
\bibitem[\protect\citeauthoryear{Yaron et al.}{2017}]{Yaron2017}
  Yaron~O. Perley~D.~A., Gal-Yam~A. et al., 2017, Nature, doi:10.1038/nphys4025
\bibitem[\protect\citeauthoryear{Yoon \& Cantiello}{2010}]{Yoon2010}
  Yoon~S.~C., Cantiello~M., 2010, ApJ, 717, L62
\end{thebibliography}
\end{document}